\documentclass[12pt]{article}

\usepackage{amsmath}
\usepackage{amssymb}
\usepackage{latexsym}
\usepackage{graphicx}
\usepackage{psfrag,fancyhdr,epsfig}

\def\beq{\begin{equation}}
\def\eeq{\end{equation}}
\def\bea{\begin{eqnarray}}
\def\eea{\end{eqnarray}}

\begin{document}

\begin{titlepage}

\hfill hep-th/0701288

\vspace*{1cm}
\begin{center}
{\bf \Large Greybody Factors for Brane Scalar Fields\\[2mm] in a
Rotating Black-Hole Background}

\bigskip \bigskip \medskip

{\bf S. Creek}$^1$, {\bf O. Efthimiou}$^2$, {\bf P. Kanti}$^{1,2}$ and
{\bf K. Tamvakis}$^2$

\bigskip
$^1$ {\it Department of Mathematical Sciences, University of Durham,\\
Science Site, South Road, Durham DH1 3LE, United Kingdom}

$^2$ {\it Division of Theoretical Physics, Department of Physics,\\
University of Ioannina, Ioannina GR-45110, Greece}

\bigskip \medskip
{\bf Abstract}
\end{center}
We study the evaporation of $(4+n)$-dimensional rotating black
holes into scalar degrees of freedom on the brane. We calculate
the corresponding absorption probabilities and cross-sections
obtaining analytic solutions in the low-energy regime, and compare
the derived analytic expressions to numerical results, with very good
agreement. We then consider the high-energy regime, construct an
analytic high-energy solution to the scalar-field equation by employing
a new method, and calculate the absorption probability and cross-section
for this energy regime, finding again a very good agreement with the
exact numerical results. We also determine the high-energy asymptotic
value of the total cross-section, and compare it to the analytic results
derived from the application of the geometrical optics limit.

\end{titlepage}

\section{Introduction}

Among the motivations for consideration of higher-dimensional theories
\cite{ADD, RS} is that the leading candidates (String Theory and variants
of it), for the unification of gravity with the rest of the fundamental
interactions at the quantum level, are all formulated in a higher-dimensional
context. In these models, gravity propagates in $D=4+n$ dimensions
({\textit{Bulk}}), while matter degrees of freedom are confined to live
on a 4-dimensional hypersurface ({\textit{Brane}}).
In models with large extra dimensions \cite{ADD},
the traditional Planck scale $M_{Pl}\sim 10^{18}\,GeV$ is only an effective
scale, related to the fundamental higher-dimensional gravity scale
$M_{*}$ through the relation $M_{Pl}^2\sim M_{*}^{n+2}R^n$, where
$R\sim (V_n)^{1/n}$ is the effective size of the $n$ extra spatial dimensions.
If $R \gg \ell_{Pl}\approx 10^{-33}\,cm$, the scale $M_{*}$ can be
substantially lower than $M_{Pl}$. In that case, trans-planckian particle
collisions could probe the strong-gravity regime and possibly produce
higher-dimensional black holes \cite{creation} centered at the brane and
extending in the bulk. For all the classical laws of black-hole physics
to still hold, the mass of the black hole $M_{BH}$ would have to be
larger than $M_{*}$. Nevertheless, the properties of these higher-dimensional
black holes would still be modified compared to their 4-dimensional analogues
\cite{ADMR, Kanti}.

If $M_{*}$ is sufficiently low, such black holes may be produced in
ground-based colliders \cite{colliders}, although their appearance in
cosmic rays is possible as well \cite{cosmic} (for 
reviews, see \cite{Kanti, reviews, Harris}). A black hole created in
such trans-planckian collisions is expected to gradually lose its angular
momentum and finally its mass through the emission of Hawking radiation
\cite{hawking}, consisting of elementary particles of a characteristic
thermal spectrum, both in the bulk and on the brane. The emitted radiation
from a higher-dimensional black hole created in trans-planckian collisions
has been studied both analytically and numerically. Until recently, the
{\it Schwarzschild} phase, along with a variety of additional
spherically-symmetric black-hole backgrounds, were the most commonly
studied cases. Those studies included the black-hole emission of lower-spin
degrees of freedom \cite{kmr1, FS, HK1, BGK, Barrau, Jung, Dai, Liu} as well
as gravitons \cite{Naylor, Park, Cardoso, Creek}, both on the brane and in the bulk.

The complexity of the gravitational background around an axially-symmetric
black hole, increased by the presence of extra dimensions, deterred many
researchers from investigating the radiation spectrum of a higher-dimensional
rotating black hole. However, during the last two years, a plethora of
studies of the Hawking radiation, emitted by such a black hole, appeared
in the literature \cite{HK2, IOP-proc, DHKW, CDKW, IOP2, Jung-super, rot-bulk,
Jung-rot}. These works offered exact numerical results that supplemented
and generalized two early analytic works focussed on the particular case
of a 5-dimensional rotating black hole \cite{Frolov2, IOP1}. Nevertheless,
to our knowledge, up to now no work has performed a complete analytic study
of the emission of Hawking radiation from a rotating black hole in an
arbitrary number of dimensions.

In the present article, we consider the evaporation of a
$(4+n)$-dimensional rotating black hole into scalar degrees of
freedom on the brane. We calculate the corresponding absorption
probabilities and cross-sections obtaining analytical solutions
in both the high and low-energy regimes. In section 2, we consider
the metric corresponding to a higher-dimensional rotating black hole,
and write down the equation for scalar fields propagating in the
projected-on-the-brane
background. In section 3, we focus on the low-energy regime, and
solve analytically the scalar-field equation employing the
{\textit{matching technique}} of combining the far-field and
near-horizon parts of the solution. Subsequently, we derive an
analytic expression for the absorption probability, and produce a
set of plots exhibiting its dependence on particle quantum numbers
and topological properties of spacetime. We proceed by deriving
corresponding numerical plots, and compare them to the analytic ones,
with excellent agreement. We finally derive a low-energy, simplified
expression for the absorption cross-section, and confirm the universal
behaviour characterising the absorption of scalar fields in this
particular energy regime. In section 4, we turn our attention to
the high-energy regime. We construct an analytic high-energy solution
to the scalar-field equation by employing a new method involving
{\textit{Kummer functions}}, and calculate the absorption probability
at this energy regime. By using our analytic results, we produce
plots which we compare to the corresponding numerical ones finding
again a very good agreement. In the same section, we determine, through
numerical integration, the exact asymptotic value of the total absorption
cross-section at the high-energy regime. We also include an analytic
treatment of the {\textit{geometrical optics limit}}, which is
expected to correspond to the high-energy asymptotic regime. We 
derive expressions for the absorption cross-section in this limit
for three distinct kinematical cases, and compare them with the exact
numerical results. Finally, in section 5, we state our conclusions.


\section{Gravitational Background and Field Equations}

If one accepts the prospect of the creation of a microscopic black
hole during a high-energy particle collision, then, due to a
generically non-vanishing value of the impact parameter between
the two particles, the emergence of a rotating black hole is the
most natural outcome. As the black hole is created in the
framework of the higher-dimensional theory, that is characterized
by a strong gravitational force, it will itself be a
higher-dimensional object, that ``feels'' the extra compact,
spacelike dimensions. Under the assumption that the black-hole
horizon $r_h$ is significantly smaller than the size of the extra
dimensions, the spacetime around it may be approximated by one
with a single timelike dimension and $(3+n)$ non-compact,
spacelike ones. A black hole living in such a background can have
in general up to $[(n+3)/2]$ angular momentum parameters. However,
here, we will be assuming that the colliding particles are
restricted to propagate on an infinitely-thin 3-brane, therefore,
they will have a non-zero impact parameter only along our brane,
and thus acquire only one non-zero angular momentum parameter
about an axis in the brane. The background around a
higher-dimensional rotating black hole with one angular momentum
parameter is given by the following Myers-Perry solution \cite{MP}
\begin{eqnarray}
&~& \hspace*{-3cm}ds^2 = \biggl(1-\frac{\mu}{\Sigma\,r^{n-1}}\biggr) dt^2 +
\frac{2 a \mu \sin^2\theta}{\Sigma\,r^{n-1}}\,dt \, d\varphi
-\frac{\Sigma}{\Delta}\,dr^2 -\Sigma\,d\theta^2 \nonumber \\[2mm]
\hspace*{2cm}
&-& \biggl(r^2+a^2+\frac{a^2 \mu \sin^2\theta}{\Sigma\,r^{n-1}}\biggr)
\sin^2\theta\,d\varphi^2 - r^2 \cos^2\theta\, d\Omega_{n}^2,
\label{rot-metric}
\end{eqnarray}
where
\begin{equation}
\Delta = r^2 + a^2 -\frac{\mu}{r^{n-1}}\,, \qquad
\Sigma=r^2 +a^2\,\cos^2\theta\,,
\label{Delta}
\end{equation}
and $d\Omega^2_n$ is the line-element on a unit $n$-sphere. The
mass and angular momentum of the black hole are then given by
\begin{equation}
M_{BH}=\frac{(n+2) A_{n+2}}{16 \pi G}\,\mu\,,  \qquad
J=\frac{2}{n+2}\,M_{BH}\,a\,, \label{def}
\end{equation}
with $G$ being the $(4+n)$-dimensional Newton's constant, and $A_{n+2}$
the area of a $(n+2)$-dimensional unit sphere given by
\begin{equation}
A_{n+2}=\frac{2 \pi^{(n+3)/2}}{\Gamma[(n+3)/2]}\,.
\end{equation}

Since the creation of the black hole depends crucially on the value of the
impact parameter between the two highly-energetic particles \cite{creation},
and that in turn defines the angular momentum of the black hole, an upper
bound can be imposed on the angular momentum parameter $a$ of the black hole
by demanding the creation of the black hole itself during the collision.
The maximum value of the impact parameter between the two particles
that can lead to the creation of a black hole is \cite{Harris}
\begin{equation}
b_\text{max}=2 \,\biggl[1+\biggl(\frac{n+2}{2}\biggr)^2\biggr]^{-\frac{1}{(n+1)}}
\mu^{\frac{1}{(n+1)}}\,,
\end{equation}
an analytic expression that is in very good agreement with the
numerical results produced in the third paper of Ref.
\cite{creation}. If we write $J=b M_{BH}/2$, for the angular
momentum of the black hole, and use the following expression for
the black-hole horizon
\begin{equation}
r_{h}^{n+1}=\frac{\mu}{1+a_*^2}\,,
\label{horizon}
\end{equation}
that follows from the equation $\Delta(r)=0$, and the second of Eqs. (\ref{def}),
we obtain
\begin{equation}
a^\text{max}_*=\frac{n+2}{2}\,.
\label{amax}
\end{equation}
In the above, we have defined, for convenience, the quantity $a_*=a/r_{h}$.
Equation (\ref{amax}), thus, puts an upper bound to the value of the
black-hole angular momentum parameter, which for $n>1$ would have been
unrestricted, contrary to the cases of $n=0$ and $n=1$, where a maximum
value of $a$ exists that guarantees the existence of a real solution
for the black-hole horizon.

In this work, we will focus on the propagation of scalar fields in the
gravitational background induced on the brane, where all ordinary particles
are assumed to live. The 4-dimensional induced background will be the
projection of the higher-dimensional one onto the brane, and its exact
expression follows by fixing the values of the additional azimuthal
angular variables -- introduced to describe the $n$ compact extra dimensions --
to $\theta_i=\pi/2$, for $i=2,...,n+1$. Then, the induced-on-the-brane
line-element takes the form
\begin{equation}
\begin{split}
ds^2=\left(1-\frac{\mu}{\Sigma\,r^{n-1}}\right)dt^2&+\frac{2 a\mu\sin^2\theta}
{\Sigma\,r^{n-1}}\,dt\,d\varphi-\frac{\Sigma}{\Delta}dr^2 \\[3mm] &\hspace*{-1cm}
-\Sigma\,d\theta^2-\left(r^2+a^2+\frac{a^2\mu\sin^2\theta}{\Sigma\,r^{n-1}}\right)
\sin^2\theta\,d\varphi^2\,.
\end{split} \label{induced}
\end{equation}
Note that although the above background is very similar to the usual
4-dimensional Kerr one, it is not exactly the same due to its explicit
dependence on the number of additional spacelike dimensions $n$. It is
this dependence that will cause brane quantities to depend on the number
of dimensions that exist transverse to the brane.

In order to study the propagation of fields in the above background,
we need to derive first their equations of motion. We assume
that the particles couple only minimally to the gravitational background
and have no other interactions, therefore, they satisfy the corresponding
free equations of motion. The latter, for particles with spin 0, 1/2 and 1,
propagating in the induced-on-the-brane gravitational background (\ref{induced}),
were derived in \cite{Kanti, IOP1}. For scalar fields, the field factorization
\begin{equation}
\phi(t,r,\theta,\varphi)= e^{-i\omega t}\,e^{i m \varphi}\,R_{\omega \ell m}(r)
\,T^{m}_{\ell}(\theta, a \omega)\,,
\end{equation}
where $T^{m}_{\ell}(\theta, a \omega)$ are the so-called spheroidal harmonics
\cite{spheroidals}, was shown to lead to the following set of decoupled radial
and angular equations
\begin{equation}
\frac{d}{dr}\biggl(\Delta\,\frac{d R_{\omega \ell m}}{dr}\biggr)+
\left(\frac{K^2}{\Delta} -\Lambda^m_{\ell}
\right)R_{\omega \ell m}=0\,, \label{radial}
\end{equation}
\smallskip
\begin{equation}
\label{angular}
\frac{1}{\sin\theta} \frac{d}{d\theta}\left(\sin\theta\,\frac{d T^m_{\ell}(\theta, a \omega)}
{d\theta}\right) + \biggl(-\frac{m^2}{\sin^2\theta}
+a^2\omega^2\cos^2\theta + E^m_{\ell}\biggr) T^m_{\ell}(\theta, a \omega)=0\,,
\end{equation}
respectively. In the above, we have defined
\begin{equation}
K=(r^2+a^2)\,\omega-am\,,  \qquad
\Lambda^m_{\ell}=E^m_{\ell}+a^2\omega^2-2am\omega\,.
\end{equation}
The angular eigenvalue $E^m_{\ell} (a\omega)$ provides a link between the angular
and radial equation. Its expression, in general, cannot be written in closed form,
however, an analytic form can be found \cite{eigenvalue} in terms of a power series
with respect to the parameter $a \omega$. We will return to this point in Section 3.

By solving Eq. (\ref{radial}), we determine the radial part of the field wavefunction,
and, subsequently, the absorption probability $|{\cal A}_{\ell,m}|^2$ for the
propagation of a scalar field in the projected-on-the-brane background. This
quantity appears in the differential emission rates for Hawking radiation emitted
by the higher-dimensional black hole on the brane. For example, the particle flux,
i.e. the number of particles emitted per unit time and unit frequency, has the form
\begin{equation}
\frac {d^{2}N}{dt  d\omega}  =
\frac {1}{2\pi} \sum _{\ell ,m}
\frac {1}{\exp\left[k/T_\text{H}\right] - 1}
|{\cal A}_{\ell,m}|^2\,;\label{flux}
\end{equation}
similar formulae may be written for the energy and angular momentum emission rates.
In the above, $k$ and $T_H$ stand for
\beq
k  \equiv \omega - m \Omega =\omega - \frac{m a}{r_h^2 +a^2}\,, \qquad \quad
T_\text{H}=\frac{(n+1)+(n-1)a_*^2}{4\pi(1+a_*^2)r_{h}}\,,
\label{k}
\eeq
with $\Omega$ the angular velocity of the black hole and $T_H$ its temperature.
The absorption probability $|{\cal A}_{\ell,m}|^2$ depends both on particle
properties (energy $\omega$, angular momentum numbers $\ell, m$, etc) and
gravitational background properties (number of extra dimensions $n$,
black-hole angular momentum parameter $a$). As a result, it modifies the
various emission rates from the ones for a blackbody. Equation (\ref{radial})
has been solved analytically only for the case of a 5-dimensional rotating
black hole (and then, only in the low-energy regime) \cite{IOP1}, and numerically
in \cite{HK2, DHKW, IOP2} for arbitrary dimensions\footnote{The angular equation
(\ref{angular}) was also solved numerically in \cite{DHKW}, and the angular
distribution of the emitted radiation was found.}. In the next sections, we
will attempt to derive analytic results, both in the low- and
high-energy regimes, for the absorption probability for scalar fields
propagating in the 4-dimensional spacetime of a brane embedded in the background
of a rotating black hole of arbitrary dimensionality.


\section{Greybody Factor in the Low-Energy Regime}

In this section, we focus on the solution for the absorption probability
in the low-energy regime. We will first derive an analytic expression
by using a well-known approximate method. We will then plot this expression
to reveal its dependence on a number of parameters, such as the angular momentum
numbers of the particle, the dimensionality of spacetime, and the angular
momentum of the black hole. It will also be directly compared with the
exact numerical results derived earlier in the literature. We will
finally derive a compact analytic expression, valid in the limit
$\omega \rightarrow 0$, and comment on the form of the corresponding
absorption cross-section and its relation to the area of the
black-hole horizon.

\subsection{Solving the Field Equation Analytically}

In what follows, we will use an approximate method and solve first the radial
equation of motion (\ref{radial}) at the two asymptotic regimes: close to the
black-hole horizon ($ r \simeq r_h$), and far away from it ($r \gg r_h$). The
two solutions will then be stretched and matched in an intermediate zone to
create a smooth analytical solution extending over the whole radial regime.

We first focus on the near-horizon regime and make the following change
of variable
\beq
r \rightarrow f(r) = \frac{\Delta(r)}{r^2+a^2} \,\,\Longrightarrow\,
\frac{df}{dr}=(1-f)\,r\,\frac{A(r)}{r^2+a^2}\,,
\eeq
where, for convenience, we have defined the function $A(r) \equiv (n+1)+(n-1)\,a^2/r^2$.
Then, Eq. (\ref{radial}), near the horizon ($ r \simeq r_h$), takes the form
\begin{equation}
f\,(1-f)\,\frac{d^2P}{df^2} + (1-D_*\,f)\,\frac{d P}{df}
+ \biggl[\,\frac{K^2_*}{A_*^2\,f (1-f)}
-\frac{\Lambda^m_\ell\,(1+a_*^2)}{A_*^2\,(1-f)}\,\biggr] P=0\,,
\label{NH-1}
\end{equation}
where now
\begin{equation}
A_*=(n+1) + (n-1)\,a_*^2\,, \qquad
K_*=(1+a_*^2)\,\omega_* - a_* m\,,
\label{K-A}
\end{equation}
with $\omega_* \equiv \omega r_h$. We have also defined the quantity
\beq
D_* \equiv 1+ \frac{n\,(1+a_*^2)}{A_*} - \frac{4 a_*^2}{A_*^2}\,.
\label{Dstar}
\eeq

By making the field redefinition $P(f)=f^\alpha (1-f)^\beta F(f)$,
Eq. (\ref{NH-1}) takes the form of a hypergeometric equation \cite{Abramowitz}:
\beq
f\,(1-f)\,\frac{d^2 F}{df^2} + [c-(1+a+b)\,f]\,\frac{d F}{df} -ab\,F=0\,,
\label{hyper}
\eeq
with
\begin{eqnarray}
a=\alpha + \beta +D_*-1\,, \qquad
b=\alpha + \beta\,, \qquad c=1 + 2 \alpha\,.
\end{eqnarray}
The power coefficients $\alpha$ and $\beta$ will be determined by two
constraints (following from the demand that the coefficient of $F(f)$ is
indeed $-ab$) that have the form of second-order algebraic equations, namely
\beq
\alpha^2 + \frac{K_*^2}{A_*^2}=0\,,
\label{alpha-eq}
\eeq
and
\beq
\beta^2 + \beta\,(D_*-2) + \frac{K_*^2}{A_*^2} -
\frac{\Lambda^m_\ell\,(1+a_*^2)}{A_*^2}=0\,. \label{beta-eq}
\eeq

The general solution of the hypergeometric equation (\ref{hyper}), combined
with the relation between $P(f)$ and $F(f)$, leads to the following
expression for the radial function $P(f)$ in the near-horizon regime:
\begin{eqnarray}
&& \hspace*{-1cm}P_{NH}(f)=A_- f^{\alpha}\,(1-f)^\beta\,F(a,b,c;f)
\nonumber \\[1mm] && \hspace*{2cm} +\,
A_+\,f^{-\alpha}\,(1-f)^\beta\,F(a-c+1,b-c+1,2-c;f)\,,
\label{NH-gen}
\end{eqnarray}
\noindent
where $A_{\pm}$ are arbitrary constants. Solving Eq. (\ref{alpha-eq}),
we obtain the solutions
\begin{equation}
\alpha_{\pm} = \pm\frac{i K_*}{A_*}\,. \label{alpha}
\end{equation}
Near the horizon, $r \rightarrow r_h$ and $f(r) \rightarrow 0$. Then,
the near-horizon solution (\ref{NH-gen}) reduces to
\beq
P_{NH}(f) \simeq A_-\,f^{\pm i K_*/A_*} + A_+\,f^{\mp i K_*/A_*}=
A_-\,e^{\pm i ky} + A_+\,e^{\mp i ky }\,,
\end{equation}
where, in the second part, we have used the definition for $k$ given in
Eq. (\ref{k}), and the tortoise-like coordinate
$y= r_h (1+a_*^2) \ln(f)/A_*$. Note, that although the coordinate
$y$ is not identical to the usual tortoise-one, defined by
$dr_*/dr=(r^2+a^2)/\Delta(r)$, used in Kerr-like backgrounds \cite{Chandra},
it holds that
\beq
\frac{dy}{dr}=\left(\frac{A}{A_*}\right)\,\frac{(r_h^2+a^2)^2}{(r^2+a^2)^2}\,
\left(\frac{r_h}{r}\right)^{n-2}\,\frac{dr_*}{dr}\,.
\eeq
Therefore, in the limit $r \rightarrow r_h$, the two become
identical, and the near-horizon asymptotic solution assumes, as
expected, the free-wave form in terms of the tortoise-coordinate
\cite{HK2,DHKW,Chandra}. Imposing the boundary condition that no
outgoing mode exists near the horizon, we are forced to set either
$A_-=0$ or $A_+=0$, depending on the choice for $\alpha$. As the
two are clearly equivalent, we choose $\alpha=\alpha_-$, and set
$A_+=0$. This brings our near-horizon solution to the final form
\beq
P_{NH}(f)=A_- f^{\alpha}\,(1-f)^\beta\,F(a,b,c;f)\,. \label{NH-final}
\eeq
We finally turn to Eq. (\ref{beta-eq}) for the $\beta$ power coefficient. This
admits the solutions
\begin{equation}
\beta_{\pm} =\frac{1}{2}\,\biggl[\,(2-D_*)\pm \sqrt{(D_*-2)^2 -
\frac{4 K_*^2}{A_*^2} +
\frac{4\Lambda^m_\ell\,(1+a_*^2)}{A_*^2}}\,\,\biggr]\,.
\label{beta}
\end{equation}
The sign appearing in front of the square root will be decided by the
criterion for the convergence of the hypergeometric function $F(a,b,c;f)$,
i.e. ${\rm Re}\,(c-a-b)>0$, which demands that we choose $\beta=\beta_{-}$.

We now turn our attention to the far-field regime. Making the assumption
that $r \gg r_h$, and keeping only the dominant terms in the expansion
in terms of $1/r$, the radial equation (\ref{radial}) takes the form
\beq
\frac{d^2P}{dr^2}+ \frac{2}{r}\frac{dP}{dr}+\left(\omega^2 -
\frac{E^m_{l}+a^2\omega^2}{r^2} \right)P(r)=0 \,,
\label{FF-eq}
\eeq
where $E^m_l$ is the angular eigenvalue. The substitution
$P(r)= \frac{1}{\sqrt{r}}\,\tilde P(r)$ brings the above into the form of
a Bessel equation for $\tilde P(r)$, and the overall solution in the far-field
limit can be written as
\beq
P_{FF}(r)=\frac{B_1}{\sqrt{r}}\,J_{\nu}\,(\omega r) +
\frac{B_2}{\sqrt{r}}\,Y_{\nu}\,(\omega r) \,,
\label{FF}
\eeq
where $J_\nu$ and $Y_\nu$ are the Bessel functions of the first and second
kind, respectively, with $\nu=\sqrt{E^m_l+a^2\omega^2+1/4}$,
and $B_{1,2}$ integration constants.

In order to construct an analytic solution extending over the whole radial
regime, we need to smoothly match the two asymptotic solutions, derived above,
in some intermediate regime. Before doing so, we first need to extrapolate
(stretch) the two solutions towards this regime. To this end, we shift the
argument of the hypergeometric function of the near-horizon solution from $f$
to $1-f$ by using the relation \cite{Abramowitz}
\bea
\hspace*{-2mm}P_{NH}(f)&=&A_- f^\alpha\,(1-f)^\beta\,\Biggl[\,
\frac{\Gamma(c)\,\Gamma(c-a-b)}
{\Gamma(c-a)\,\Gamma(c-b)}\,F(a,b,a+b-c+1;1-f) \nonumber \\[2mm]
&+& (1-f)^{c-a-b}\,\frac{\Gamma(c)\,\Gamma(a+b-c)}
{\Gamma(a)\,\Gamma(b)}\,F(c-a,c-b,c-a-b+1;1-f)\,\Biggr]. \label{NH-shifted}
\eea
The function $f(r)$ may be alternatively written as
\beq
f(r)=1-\frac{\mu}{r^{n-1}}\,\frac{1}{r^2+a^2}=
1-\biggl(\frac{r_h}{r}\biggr)^{n-1}\,\frac{(1+a_*^2)}
{(r/r_h)^2+a_*^2}\,,
\label{f-far}
\eeq
where we have used the horizon equation,  $\Delta(r_h)=0$, in order to eliminate
$\mu$ from the above relation. In the limit $r \gg r_h$, the $(r/r_h)^2$ in
the denominator of the second term above is dominant, and the whole expression
goes to unity for $n \geq 0$. Thus, in the limit $f \rightarrow 1$,
the near-horizon solution (\ref{NH-shifted}) takes the form
\beq
P_{NH}(r) \simeq A_1\, r^{\,-(n+1)\,\beta} + A_2\,r^{\,(n+1)\,(\beta + D_*-2)}\,,
\label{NH-stretched}
\eeq
with
\bea
A_1 &=& A_- \left[(1+a_*^2)\,r_h^{n+1}\right]^\beta \,
\frac{\Gamma(c)\Gamma(c-a-b)}{\Gamma(c-a)\Gamma(c-b)} \nonumber\\[1mm]
A_2 &=& A_- \left[(1+a_*^2)\,r_h^{n+1}\right]^{-(\beta + D_*-2)} \,
\frac{\Gamma(c)\Gamma(a+b-c)}{\Gamma(a)\Gamma(b)} \,.
\eea

Next we need to stretch the far-field asymptotic solution (\ref{FF}) towards
small values of the radial coordinate. Then, in the limit $\omega r \rightarrow 0$,
we find
\bea
&~& \hspace*{-5.0cm}P_{FF}(r) \simeq \frac{B_1\left(\frac{\omega r}{2}\right)^
{\sqrt{E^m_l+a^2\omega^2+1/4}}}{\sqrt{r} \,\Gamma\left(\sqrt{E^m_l+a^2\omega^2+1/4}+1\right)}
\nonumber \\[2mm]
\hspace*{3.5cm}
&-& \frac{B_2}{\pi \, \sqrt{r}}\,\Gamma\left(\sqrt{E^m_l+a^2\omega^2+1/4}\right)
\left(\frac{\omega r}{2}\right)^{-\sqrt{E^m_l+a^2\omega^2+1/4}} \,.
\label{FF-stretched}
\eea

We notice that both `stretched' asymptotic solutions have reduced to power-law
expressions in terms of the radial coordinate $r$, however, the different
power coefficients prevent the exact matching. In order to overcome this obstacle,
we will expand these power coefficients in the limits $(\omega r_h)^2 \ll 1$ and
$(a/r_h)^2 \ll 1$. It is these approximations that will limit the validity
of our result for the absorption probability to the low-energy and
low-angular-momentum regime. Note, however, that in order to improve the
accuracy of our result, no approximation will be made in the arguments
of the Gamma functions involved in Eqs. (\ref{NH-stretched}) and
(\ref{FF-stretched}). In order to follow the aforementioned line of action,
we need to know the analytic expression for the angular eigenvalue $E^m_l$.
According to \cite{eigenvalue}, this may be written as a power series in
terms of the parameter $(a\omega)$, namely
\beq
E^m_l=\sum_{n=0}^{\infty}\,f_n^{lm}\,(a \omega)^n\,.
\eeq
For the purposes of our analysis, we have calculated the coefficients
$f_n^{lm}$ up to fifth order, and found the result
\bea
E^m_l &=& l\,(l+1)+ (a\omega)^2\,\frac{[2m^2-2l\,(l+1)+1]}{(2l-1)\,(2l+3)}
\nonumber \\[1mm] &+&
(a\omega)^4\,\left\{\frac{2\,[-3+17 l\,(l+1) +l^2 (l+1)^2(2l-3)\,(2l+5)]}
{(2l-3)\,(2l+5)\,(2l+3)^3(2l-1)^3} \right. \nonumber\\[1mm]
&+& \left. \frac{4m^2}{(2l-1)^2(2l+3)^2}\,
\left[\frac{1}{(2l-1)\,(2l+3)} -\frac{3l\,(l+1)}{(2l-3)\,(2l+5)}\right] \right.
\nonumber \\[1mm] &+& \left.
\frac{2m^4\,[48 +5 (2l-1)\,(2l+3)]}{(2l-3)\,(2l+5)\,(2l-1)^3(2l+3)^3}\right\} + ....\,,
\eea
with $f_1^{lm}=f_3^{lm}=f_5^{lm}=0$. The above form will be used at every
place where $E^m_l$ appears in Eqs. (\ref{NH-stretched}) and (\ref{FF-stretched}).
The only exception will be in the power coefficients, where terms of order
$(a\omega)^2$, or higher, will be ignored. Following these assumptions, the two
power coefficients in Eq. (\ref{NH-stretched}) reduce to
\bea
-(n+1)\,\beta &\simeq& l + {\cal O}(\omega_*^2, a_*^2, a_*\omega_*)\,, \nonumber \\[1mm]
(n+1)\,(\beta+D_*-2) &\simeq& -(l+1) + {\cal O}(\omega_*^2, a_*^2, a_*\omega_*)\,.
\eea
while the power coefficient in Eq. (\ref{FF-stretched}) takes the form
\beq
\sqrt{E^m_l+a^2\omega^2+1/4} \simeq \left(l+\frac{1}{2}\right) +
{\cal O}(a_*^2\omega_*^2)\,.
\eeq
By using the above results, one can easily show that both Eqs. (\ref{NH-stretched})
and (\ref{FF-stretched}) reduce to power-law expressions with the same power
coefficients, $r^l$ and $r^{-(l+1)}$. By matching the corresponding coefficients,
we determine the integration constants $B_{1,2}$ in terms of the other parameters
of the theory. Their ratio, which, as we shall see, appears in the expression of the
absorption probability $|{\cal A}_{l,m}|^2$, is found to be
\bea
&~& \hspace*{-1.8cm}B \equiv \frac{B_1}{B_2} = -\frac{1}{\pi}
\left(\frac{2}{\omega r_h\,(1+a_*^2)^\frac{1}{n+1}}\right)^{2l+1}
\sqrt{E^m_l+a^2\omega^2+1/4} \nonumber \\[2mm]
\hspace*{0.4cm}
&& \times\,\frac{\Gamma^2\left(\sqrt{E^m_l+a^2\omega^2+1/4}\right)
\Gamma(\alpha+\beta + D_* -1)\,\Gamma(\alpha+\beta)\,
\Gamma(2-2\beta - D_*)}{\Gamma(2\beta + D_*-2)\,
\Gamma(2+\alpha -\beta - D_*)\,\Gamma(1+\alpha-\beta)} \,. \label{Beq}
\eea

The last step in our calculation of the absorption probability involves the
expansion of the far-field solution (\ref{FF}) to infinity ($r \rightarrow \infty$).
This leads to
\bea
P_{FF}(r) &\simeq& \frac{1}{\sqrt{2\pi\omega}}\left[\frac{(B_1+iB_2)}{r}\,
e^{-i\,\left(\omega r - \frac{\pi}{2}\,\nu - \frac{\pi}{4}\right)}
+ \frac{(B_1-iB_2)}{r}\,e^{i\,\left(\omega r - \frac{\pi}{2}\,\nu -
\frac{\pi}{4}\right)} \right]\nonumber \\
&=& A_{in}^{(\infty)}\,\frac{e^{-i\omega r}}{r} +
A_{out}^{(\infty)}\,\frac{e^{i\omega r}}{r} \,.
\eea
As expected, at large distances
from the black hole, where the effect of the angular momentum parameter $a$
is almost negligible, the solution for the scalar field reduces to a spherical wave,
as in the Schwarzschild case \cite{kmr1, HK2, DHKW}. Then, the
absorption probability is defined through the expression
\bea
\left|{\cal A}_{l,m}\right|^2 &=& 1-\left|\frac{A_{out}^{(\infty)}}{A_{in}^{(\infty)}}\right|^2
= 1-\left|\frac{B_1-iB_2}{B_1+iB_2}\right|^2 \nonumber \\
&=& 1-\left|\frac{B-i}{B+i}\right|^2 = \frac{2i\left(B^*-B\right)}{B B^*
+ i\left(B^*-B\right)+1}\,. \label{Absorption}
\eea
The above result, together with the expression for $B$ given in
Eq. (\ref{Beq}), is our main analytic result for the absorption probability
for scalar fields valid in the low-energy and low-angular-momentum regime.
It can be easily checked that it reduces smoothly to the corresponding result
for a scalar field propagating in a Schwarzschild-like, higher-dimensional
background projected onto the brane \cite{kmr1}, if we set $a=0$.


\subsection{A Comparison with the Exact Numerical Solution}

In this section, we proceed to study in detail the properties of the absorption
probability $|{\cal A}_{l,m}|^2$, as this was derived above by using purely
analytic arguments. To this end, we will plot our main result, given by
Eqs. (\ref{Absorption}) and (\ref{Beq}), as a function of the energy parameter
$\omega r_h$ and for a variety of values of the other parameters of the theory,
namely the angular momentum numbers $(l,m)$ of the scalar particle, and
the topological parameters $(a_*, n)$ of the spacetime. At the same time,
a direct comparison of our low-energy analytic result to the exact numerical
value for $|{\cal A}_{l,m}|^2$ -- derived in \cite{Harris, HK2, DHKW} for
the purpose of computing the Hawking radiation emission spectra -- will be
performed in order to examine the range of validity of our approximations.

\begin{figure}
  \begin{center}
 \includegraphics  [width = 0.70 \textwidth] {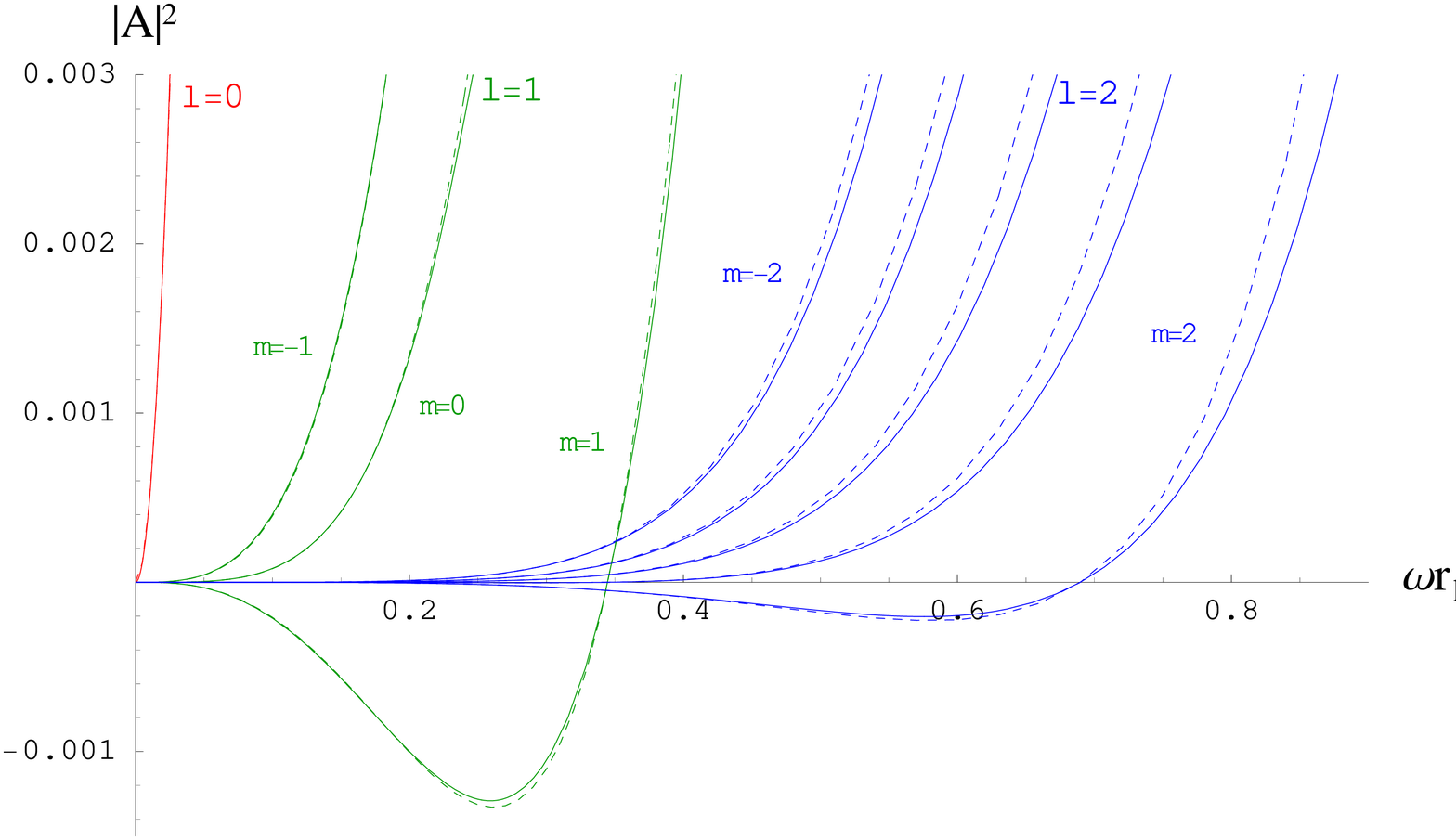}
    \caption{Absorption probability $|{\cal A}_{l,m}|^2$ for brane scalar particles,
    for $n=2$, $a_*=0.4$, and for the modes (from left to right) $(l=m=0)$,
    $(l=1, m=-1,0,1)$, and ($l=2$, $m=-2,-1,0,1,2$). The solid lines correspond
    to our analytic results, and the dashed lines to the exact numerical ones.}
    \label{figl0123m}
  \end{center}
\end{figure}

In Fig. \ref{figl0123m}, we plot the absorption probability $|{\cal A}_{l,m}|^2$
for scalar particles, for fixed angular momentum of the black hole ($a_*=0.4$)
and number of extra dimensions $(n=2)$, and for a variety of modes with
different angular momentum numbers $(l,m)$. Throughout our paper, our
analytic results will be plotted by using solid lines while the exact
numerical results, reproduced from \cite{Harris, HK2, DHKW}, will be denoted
by dashed lines. In Fig. \ref{figl0123m}, both sets of lines are shown
for all modes, and the agreement between them at low energy is
indeed remarkable. Although a small deviation appears when the energy
parameter is taken beyond the low-energy regime, the qualitative agreement
between the two sets of results remains excellent.

Focusing now on the dependence of the absorption probability on particle
parameters, we observe that, similarly to the
Schwarzschild case \cite{kmr1}, it is again the lowest partial
wave, with $l=0$, that dominates in the low-energy regime, with all higher
modes increasingly suppressed. This behaviour is valid for all values of
$a_*$ and $n$, as long as attention is focused on the low-energy regime.
Looking next at the relative behaviour of modes with the same angular
momentum number $l$ but different number $m$, we easily note that the
modes with $m<0$ are the dominant ones with increasing $m$ causing suppression.
We also observe that, for modes with $m\leq0$, the absorption probability
always remains positive, while, for modes with $m>0$, a negative-valued region
for $|{\cal A}_{l,m}|^2$ always appears in the low-energy regime. The
latter effect is due to the so-called superradiance \cite{super} -- the enhancement
of the amplitude of an incoming wave by a rotating black hole that results in a
reflection probability larger than unity, and thus a negative absorption
probability according to $|{\cal A}_{l,m}|^2 =1- |{\cal R}_{l,m}|^2$. We will say
a little more on the origin of this effect in the following subsection.

\begin{figure}
  \begin{center}
  \mbox{\includegraphics[width = 0.5 \textwidth] {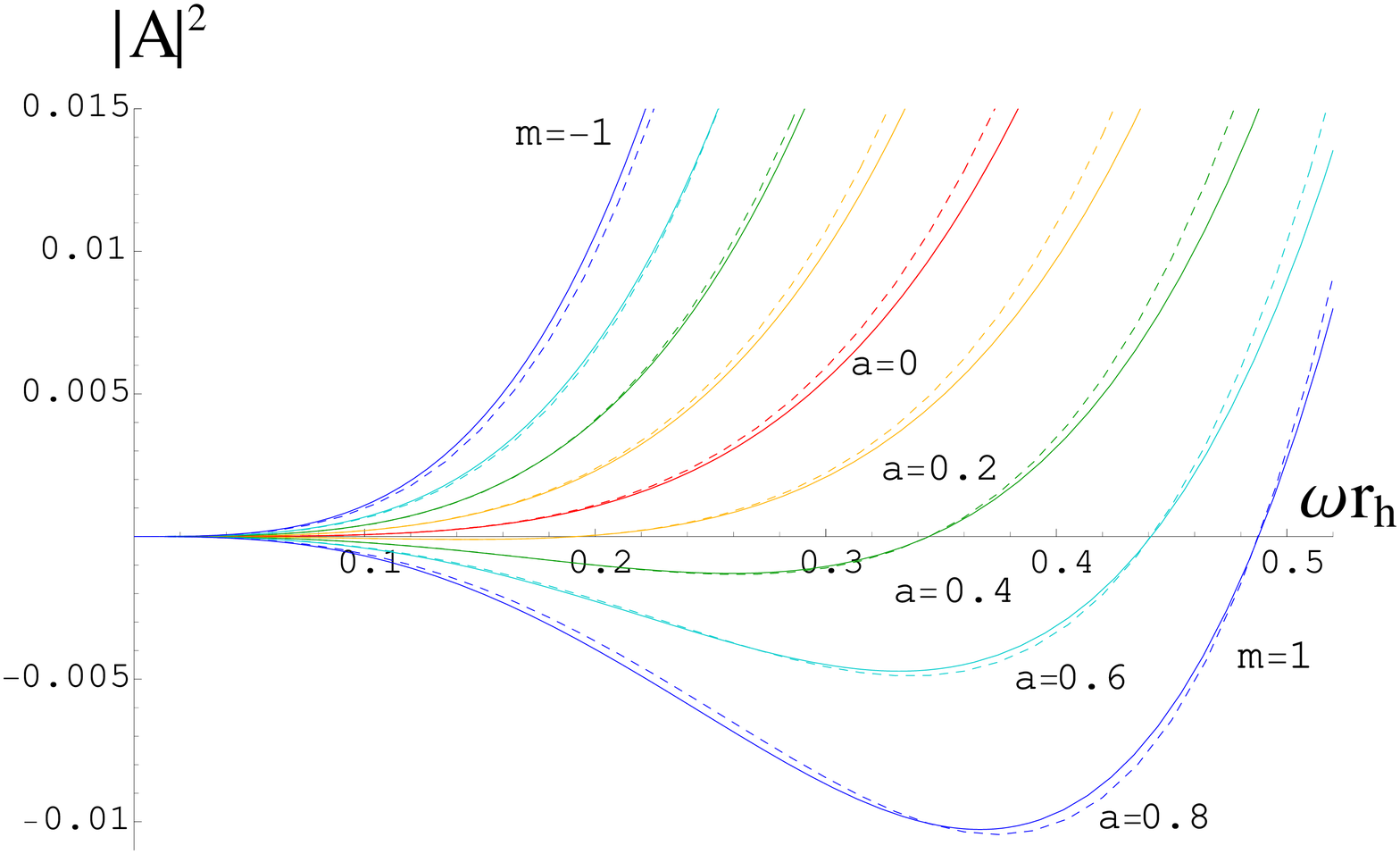}}
\hspace*{-0.3cm}
{\includegraphics[width = 0.5 \textwidth] {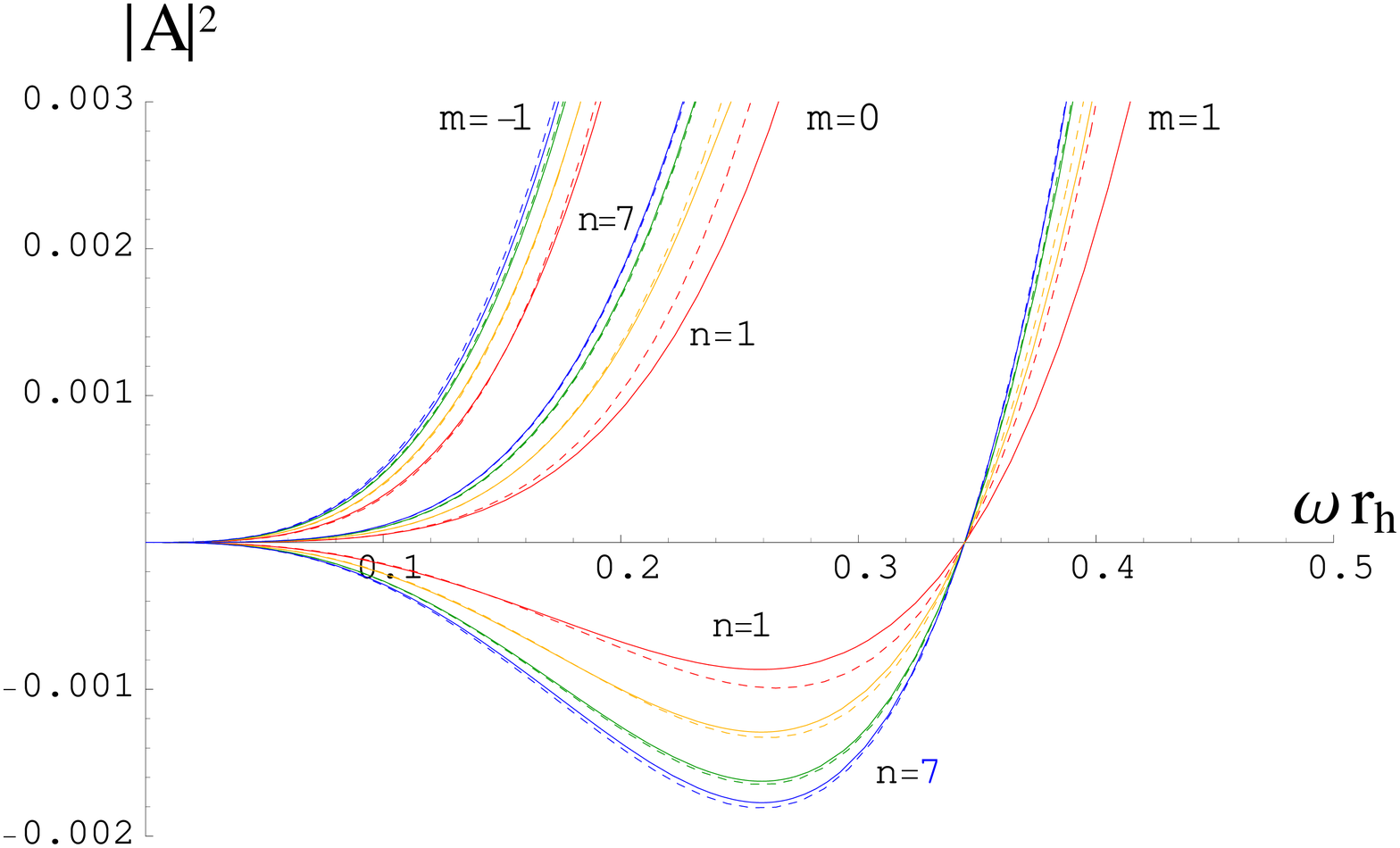}}
    \caption{Absorption probability $|{\cal A}_{l,m}|^2$ for brane scalar particles,
   {\bf (a)} for the modes $(l=1, m=-1,1)$, for $n=2$ and $a_*=(0,0.2, 0.4, 0.6, 0.8)$,
   and {\bf (b)} for the modes $(l=1, m=-1,0,1)$, for $a_*=0.4$ and $n=(1,2,4,7)$.
   Again, the solid lines correspond to our analytic results, and the dashed lines
   to the exact numerical ones.}
    \label{fig-an}
  \end{center}
\end{figure}

We now turn to the dependence of the absorption probability on the parameters
of spacetime. In Fig. \ref{fig-an}, we depict the behaviour of $|{\cal A}_{l,m}|^2$
in terms of the angular momentum parameter $a_*$ and number of
extra dimensions $n$.
We have chosen the indicative case of $l=1$, and, in Fig. \ref{fig-an}(a),
depict the behaviour of the $m=-1$ (from central to left) and $m=1$
(from central to right) modes, for a range of values of $a_*$. Then, in
Fig. \ref{fig-an}(b), we present all three modes, with $m=0, \pm 1$, for
various values of $n$. As before, both the analytic,
low-energy results as well as the exact numerical ones are shown. Again,
the agreement between the two sets of results in the low-energy regime
is remarkably good. The alert reader may note that, in general, the
agreement between the two sets is improving as $n$ increases:
this is due to the fact that, according to Eq. (\ref{f-far}), an increase in
the number of extra dimensions improves the accuracy of the assumed
behaviour of the function $f(r)$ at infinity, and consequently our approximation.
In addition, terms that have been neglected during the matching of the two
asymptotic solutions under the low-energy and low-$a_*$ assumption, such
as the $K_*^2/A_*^2$ in Eq. (\ref{beta}), become even smaller for large
values of $n$, thus improving the accuracy of our analysis.

According to Fig. \ref{fig-an}(a), for fixed $n$, the non-superradiant modes,
with $m \leq 0$ (although not shown, the $m=0$ mode exhibits a behaviour
similar to the $m=-1$), are enhanced with the angular momentum of the
black hole in the low-energy regime, while the superradiant modes, with
$m>0$, are suppressed both in the superradiant and non-superradiant energy regime.
Turning to Fig. \ref{fig-an}(b), we note that, for fixed $a_*$, an increase
in the number of extra dimensions $n$ also leads to an enhancement
of the value of the absorption probability for the non-superradiant modes.
For the superradiant ones, the behaviour of $|{\cal A}_{l,m}|^2$ depends
on the energy regime we are looking at: while it is suppressed in the
superradiant regime, it is enhanced in the non-superradiant one. From
both figures, it becomes obvious that the superradiant effect becomes
more important as either $a_*$ or $n$ increases. This enhancement, for
brane-localised scalar particles, both in terms of the angular momentum
of the black hole and the dimensionality of spacetime was also noted
in the literature \cite{HK2, IOP-proc, Jung-super}.


\subsection{The Low-Energy Asymptotic Limit of the Cross-Section}

Our last task regarding the behaviour of the absorption probability
$|{\cal A}_{l,m}|^2$ in the low-energy regime will be the derivation, from
Eqs. (\ref{Beq}) and (\ref{Absorption}), of a compact analytic expression
valid in the limit $\omega \rightarrow 0$. This simplified analytic expression
will be used to explain some of the features discussed in the previous
subsection. In addition, from this, the asymptotic
low-energy value of the corresponding absorption cross-section for
scalar fields in the background of a projected-on-the-brane rotating
black hole will also be determined.

We start our analysis by noticing that, according to Eq. (\ref{Beq}), in the
limit $\omega \rightarrow 0$, $B \sim \omega^{-(2l+1)}$ and, therefore,
$B B^* \gg i(B^*-B) \gg 1$. Then, Eq. (\ref{Absorption}) simplifies to
\beq
\left|{\cal A}_{l,m}\right|^2 \simeq  \frac{2i (B^*-B)}{B B^*}=
2i\left(\frac{1}{B} - \frac{1}{B^*}\right)\,.
\eeq
Substituting for $B$ using Eq. (\ref{Beq}), and the fact that $\alpha$
is purely imaginary, yields
\bea
\left|{\cal A}_{l,m}\right|^2 &=& \frac{-2i\pi\,\left(\omega r_h/2\right)^{2l+1}}
{(l+\frac{1}{2})\,\Gamma^2(l+\frac{1}{2})} \frac{\Gamma(2\beta +D_*-2)}
{(1+a_*^2)^{-\frac{2l+1}{n+1}}\,\Gamma(2-2\beta -D_*)} \quad \times  \nonumber\\
&&\frac{1}{\Gamma(\alpha+\beta+D_*-1)\,\Gamma(-\alpha+\beta+D_*-1)\,\Gamma(\alpha+\beta)
\,\Gamma(-\alpha+\beta)} \quad \times \nonumber\\[1mm]
&&\left[\vphantom{\frac{1}{1}}\Gamma(2+\alpha-\beta-D_*)\,\Gamma(-\alpha+\beta+D_*-1)\,
\Gamma(1+\alpha-\beta)\,\Gamma(-\alpha+\beta)-\right. \nonumber\\
&&\Gamma(2-\alpha-\beta-D_*)\,\Gamma(\alpha+\beta+D_*-1)\,\Gamma(1-\alpha-\beta)\,
\Gamma(\alpha+\beta)\left. \vphantom{\frac{1}{1}}\right]\label{abs-1} \\[1mm]
&=& \Sigma_1 \times \Sigma_2 \times \Sigma_3 \,, \nonumber
\eea
where $\Sigma_1$, $\Sigma_2$ and $\Sigma_3$ are defined by the quantities
on each of the three lines above. Focusing first our attention to $\Sigma_3$,
and using the Gamma function relation $\Gamma(z)\Gamma(1-z) = \pi/\sin \pi z$
\cite{Abramowitz}, this can be written as
\bea
\Sigma_3 = \frac{-\pi^2\,\sin(2\pi\alpha)\,\sin \pi(2\beta+D_*)}
{\sin\pi(\alpha+\beta+D_*)\,\sin\pi(-\alpha+\beta+D_*)\,
\sin\pi(\alpha+\beta)\,\sin\pi(-\alpha+\beta)}\,.
\eea
From the factor $\omega^{2l+1}$ in Eq. (\ref{abs-1}), it becomes clear that
the expression for the absorption probability at the very low-energy regime is
dominated by the lowest partial waves, a property that is in accordance with the
results presented in the previous subsection. Then, assuming that $m$ is small
and $a_* <1$, the limit $\omega \rightarrow 0$ is equivalent to
$\alpha \rightarrow 0$. Expanding terms in $\Sigma_3$ and $\Sigma_2$ to
linear order in $\alpha$ gives
\bea
\Sigma_3 = -\frac{2\pi^3\alpha\,\sin \pi(2\beta+D_*)}{\sin^2\pi(\beta+D_*)\,\sin^2\pi\beta}\,,
\qquad \qquad \Sigma_2 = \frac{1}{\Gamma(\beta+D_*-1)^2\,\Gamma(\beta)^2}\,.
\eea
Using the additional Gamma function relation $\Gamma(z)\Gamma(-z) = -\pi/z \sin \pi z$
allows the overall expression for the low-energy limit of the absorption probability
to be written as
\beq
\left|{\cal A}_{l,m}\right|^2 =\frac{4\pi \left(\omega r_h/2\right)^{2l+1} K_*
\,\sin^2\pi(2\beta+D_*)\,\Gamma^2(2\beta +D_*-2)\,\Gamma^2(1-\beta)\,(2-D_*-2\beta)}
{A_*\,(1+a_*^2)^{-\frac{2l+1}{n+1}}\,(l+\frac{1}{2})\,\Gamma^2(l+\frac{1}{2})\,
\Gamma^2(\beta+D_*-1)\,\sin^2\pi(\beta+D_*)}\,. \label{abs-2}
\eeq
In the above, we have also used the definition $\alpha \equiv -iK_*/A_*$, where,
from Eq. (\ref{K-A}),
\beq
K_*= (1+a_*^2)\,\omega_* - a_*\,m = r_h (1+a_*^2)\,(\omega -m\,\Omega)\,.
\eeq
By using Eq. (\ref{beta}), one may easily conclude that the quantity
$(2-D_*-2\beta)$ is always positive, while the same also holds for
$A_* \equiv (n+1) + (n-1) a_*^2$, for all values for $a_*$ and $n>0$.
Therefore, the overall sign of $\left|{\cal A}_{l,m}\right|^2$ is
determined by the sign of $K_*$, or equivalently of $(\omega-m\,\Omega)$.
A negative sign for the latter combination arises only for $m>0$, and
denotes the occurrence of superradiance, with $\left|{\cal A}_{l,m}\right|^2$
acquiring a negative value. The superradiance domain arises in the low-energy
regime and extends over the range of values $0<\omega<\omega_s \equiv m\,\Omega$.
The larger the value of the angular momentum parameter $a_*$, the larger the
rotation velocity $\Omega$ of the black hole, and thus the more extended the
superradiance regime becomes. This is indeed in agreement with the behaviour found
in the previous subsection.

In what follows, we focus on the dominant $s$-wave with $l=m=0$. As we will
see, this will be the only partial wave with a non-vanishing low-energy
asymptotic value of the absorption cross-section. In order to simplify further
Eq. (\ref{abs-2}), we need also to expand the expression of $\beta$ in the
limit $\omega \rightarrow 0$. It is easy to see that, in this limit,
$\beta = 0 + {\cal O}(\omega^2)$, which then allows us to write
\beq
\left|{\cal A}_{\,0}\right|^2 = \frac{4\,(\omega r_h)^2\,(1+a_*^2)}
{A_*\,(1+a_*^2)^{-1/(n+1)}\,(2-D_*)} + ... \,.
\eeq
The corresponding absorption cross-section for the dominant partial wave
is then given by \cite{cross-section}
\beq
\sigma_{\,0} = \frac{\pi}{\omega^2}\,\left|{\cal A}_{\,0}\right|^2=
4\pi\,(r_h^2+a^2)\,\frac{(1+a_*^2)^{1/(n+1)}}
{\left[(n+1)+(n-1)\,a_*^2\right]\,(2-D_*)}+ ... \,. \label{lowsigma}
\eeq
According to the above result, the absorption cross-section for the lowest mode
$l=0$ reduces to a non-vanishing asymptotic value, as $\omega \rightarrow 0$.
We may easily see, from Eq. (\ref{abs-2}), that the low-energy behaviour of
the absorption probability for any higher partial mode will be governed
by the factor $\omega^{2l+2}$, thus leading to an absorption cross-section
proportional to $\omega^{2l}$. Therefore, for all partial waves with $l\neq 0$,
the partial cross-section goes to zero, as $\omega \rightarrow 0$.

\begin{figure}
  \begin{center}
    \includegraphics  [width = 0.7 \textwidth] {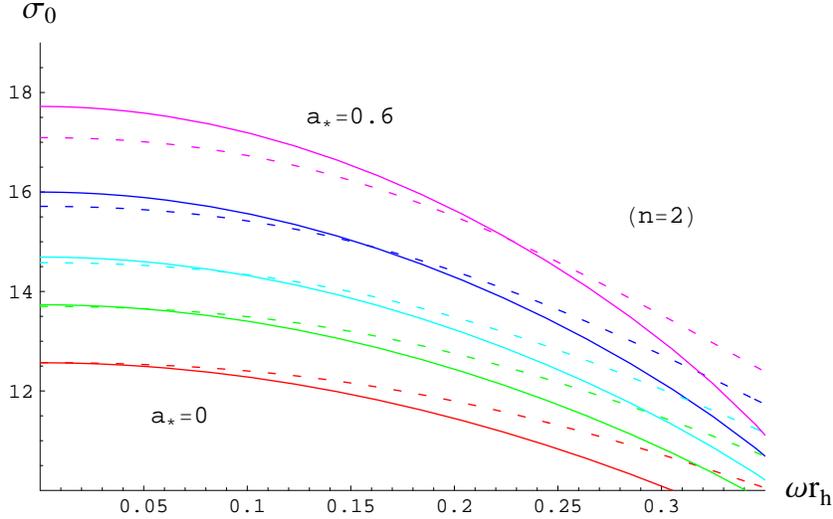}
    \caption{Absorption cross-section $\sigma_{\,0}$ (in units of $r_h^2$)
    for the lowest scalar mode
    $l=0$, for $n=2$ and $a_*$ ranging between 0 and 0.6. As before, the solid
    lines correspond to our analytic results, and the dashed lines
    to the exact numerical ones.}
    \label{cross}
  \end{center}
\end{figure}

In the case of scalar particles propagating in a Schwarzschild-like
projected-on-the-brane line-element, the low-energy asymptotic value
of the absorption cross-section of the lowest, dominant partial wave
was shown to be equal to the horizon area of the 4-dimensional black hole,
$4\pi r_h^2$, regardless of the number of extra dimensions \cite{kmr1, HK1}.
We would like to demonstrate that a similar relation holds in the case of
an axially-symmetric brane background. According to Eq. (\ref{lowsigma}),
$\sigma_{\,0}$ is indeed proportional to the horizon area of the four-dimensional
rotating black hole $4\pi(r_h^2+a^2)$, however, the relation involves a
multiplicative factor which is both ($a_*,n$)-dependent. In Fig. \ref{cross},
we plot $\sigma_{\,0}$, for fixed $n$ and various values of $a_*$, by using both
our analytic expression and the exact numerical result. The latter set of
results reveal that the asymptotic low-energy cross-section is indeed equal to
the horizon area of the black hole, regardless of the number of extra dimensions.
For small values of $a_*$, the
multiplicative factor appearing in Eq. (\ref{lowsigma}) is very close to
unity and our analytic expression closely reproduces the exact numerical one.
As $a_*$ increases though, the range of validity of our approximation is
exceeded and a deviation starts appearing, as expected. If we keep $a_*$
fixed and vary $n$ instead, a similar behaviour appears, with values of
$n\leq 1$ leading, in general, to a value smaller than the exact result,
and values of $n \ge 2$ to a value larger than the exact result.
The magnitude of the deviation depends again on the value of $a_*$.
The above results demonstrate the universal behaviour for the lowest
partial mode of a scalar particle according to which its partial cross-section
equals the area of the black-hole horizon in the low-energy regime.
This result holds not only for a spherically-symmetric brane line-element
but also for an axially-symmetric one, and is independent of the number
of the transverse-to-the-brane spacelike dimensions. The latter result
was also reproduced for the particular case of a 5-dimensional bulk in
\cite{Jung-rot}.


\section{Greybody Factor in the High-Energy Regime}

In this section, we turn our attention to the high-energy regime, and
present a way to compute the absorption probability in this region by
using again the matching technique described in section 3. For this,
we are going to use the near-horizon solution we have already computed
in Section 3.1, and we shall construct an approximate, high-energy,
far-field solution that will allow us to do the matching solely in
the high-energy regime. Next, we are going to compare the absorption
coefficient produced this way with the exact numerical results of
the literature. Finally, generalizing well-known results for Schwarzschild
black holes, the geometrical optics limit value of the
absorption cross-section will be computed, and its connection to the
high-energy asymptotic value found by numerical analysis will be investigated.

\subsection{Analytic Construction of the Solution}

In order to construct a solution over the whole radial domain that will be
valid in the high-energy regime, it is necessary to do the matching of the
stretched near-horizon and far-field asymptotic solutions without
resorting to the low-energy approximation $(\omega r_h \ll 1)$
employed in section 3.1. To this end, we will try to find a
new far-field asymptotic solution that will satisfy the field
equation only in the high-energy limit. Moreover, the exact form
of this solution should be such that, when stretched towards
small values of the radial coordinate, it reduces to a power-law
expression with identical power coefficients to the ones appearing
in the stretched near-horizon solution, thus allowing for a perfect
matching in the intermediate regime. We remind the reader that the
stretched near-horizon solution, (Eq. \ref{NH-stretched}), was found
to be  of the form
\beq  P_{NH}  \simeq A_1\,r^{\,- (n
+ 1)\beta }  + {\rm A}_2\,r^{\,(n + 1)(\beta  + D_*  - 2)}  \equiv
A_1\,r^{\,\beta _1 }  + A_2\,r^{\,\beta _2 }\,. \label{pffhe}
\eeq

The differential equation that our far-field solution needs to
satisfy is Eq. (\ref{radial}) in the limit $r \gg r_h$, or equivalently
\beq
\frac{d^2P_{FF}}{dz^2}+ \frac{2z}{z^2+a_1^2}\,\frac{dP_{FF}}{dz}+\left(1 -
\frac{E^m_{l}+a_1^2}{z^2+a_1^2} \right)P_{FF}(z)=0 \,,\label{FF-eqz}
\eeq
where we have made the change of variable $z=\omega r$, and defined for
convenience $a_1=a \omega$. Let us consider
the following trial, special solution to the above equation
\beq P_{FF}  = e^{ - i\omega r} r^{\,\beta _1} M\left(1+\beta_1,
2+2 \beta_1; 2i\omega r\right)\,, \label{ksol1} \eeq
where $M(\hat a, \hat b; w)$ is the first Kummer function \cite{Abramowitz}.
Solving Eq. (\ref{ksol1}) for $M(\hat a, \hat b;w)$ and substituting in the
confluent hypergeometric equation, that the Kummer functions satisfy,
\beq
\frac{d^2M}{dw^2}+ (\hat b-w)\,\frac{dM}{dw}-\hat a\,M(w)=0 \,,
\eeq
with $\hat a=1+\beta_1$, $\hat b=2+2\beta_1$ and $w=2i\omega r$, we finally
obtain the equation
\beq \frac{d^2 P_{FF}}{dz^2 } + \frac{2}{z}\,\frac{dP_{FF}}{dz}
+ \left( 1  - \frac{\beta _1\,(\beta _1  + 1)}{z^2 } \right)P_{FF}  = 0\,.
\label{odeFFz} \eeq
Subtracting Eq. (\ref{odeFFz}) from Eq. (\ref{FF-eqz}), we find
$$
2\,\frac{dP_{FF}}{dz} \left( \frac{z}{z^2  + a_1 ^2 } - \frac{1}{z} \right)
+ P_{FF} \left(\frac{\beta _1 (\beta _1  + 1)}{z^2 } -
\frac{E_\ell ^m +a_1^2}{z^2  + a_1 ^2 } \right)=0+ {\cal{O}}(1/z^2)\,.$$
From the above, we may conclude that, for large $z=\omega r$, the constructed
solution (\ref{ksol1}) satisfies the far-field ($r \gg r_h$) equation (\ref{FF-eqz})
up to second order in $(1/z)$. Therefore, in the limit $\omega \rightarrow \infty$,
our trial solution (\ref{ksol1}) is indeed a good approximation to the exact
solution of the field equation in the far-field regime.

By following the same method, a second special solution to Eq. (\ref{FF-eqz})
may be constructed that has the form
\beq P_{FF} = e^{ - i\omega r}\,r^{\,- \beta _2   - 1} \,U\left( { - \beta
_2, - 2\beta _2; 2i\omega r} \right)\,, \label{sol2} \eeq
where $U(\hat a, \hat b;w)$ is the second Kummer function.
The second special solution follows from the first one under the
replacement $\beta_1 \rightarrow -1-\beta_2$, which preserves the
structure of Eq. (\ref{odeFFz}). Therefore, this solution, too,
is a good approximation to the exact solution of Eq. (\ref{FF-eqz})
in the limit $\omega \rightarrow \infty$. Therefore, the total
solution in the far-field domain, valid only in the high-energy
regime, takes the form
\beq P_{FF}  = B_+ \,e^{- i\omega r} r^{\,\beta _1} M(\beta _1  + 1,
2\beta _1  + 2, 2i\omega r) + B_- \,e^{ - i\omega r} r^{\,- \beta _2- 1}
\,U( - \beta _2, - 2\beta _2, 2i\omega r)\,.
\label{heff} \eeq
The Kummer functions have been used before in the construction of the
far-field solution of a general spin-$s$ field \cite{Kanti, kmr1},
however, the projected-on-the-brane background in that case was that
of a spherically-symmetric Schwarzschild-like one. We may easily check
that, in the limit $a \rightarrow 0$, the above solution reduces to
the one for a scalar field propagating in the far-field domain of a
spherically-symmetric brane black-hole background.

As an additional independent check of the above analysis, one may observe
that, under the demand that Eq. (\ref{odeFFz}) matches exactly Eq. (\ref{FF-eqz}),
we are led to
\beq
\beta_1=-\frac{1}{2} + \sqrt{E^m_l+a^2\omega^2+1/4} = -\frac{1}{2} + \nu\,.
\eeq
But then, since $\hat a=\nu +1/2$ and $\hat b=2 \nu+1$, the first Kummer function
reduces to $M(\nu+1/2, 2\nu+1; 2i\omega r) \sim e^{i\omega r}\,r^{-\nu}
J_\nu(\omega r)$ \cite{Abramowitz}. By using this result -- as well as a similar one
for the second Kummer function $U$ -- in conjunction with the relations
(\ref{ksol1}) and (\ref{sol2}), we may see that the far-field solution (\ref{FF}),
found in section 3.1, is duly restored as expected.

Nevertheless, in this section, the approximate solution (\ref{heff}) will be
used instead, since, as was mentioned earlier, its use
will allow us to achieve a perfect matching between the stretched near-horizon
and far-field solutions -- a feature that, as we saw, was not possible when the
asymptotic solution (\ref{FF}) was used instead. To this end, we stretch
Eq. (\ref{heff}) to small values of $\omega r$ \cite{Abramowitz} to obtain
\beq
P_{FF}  = B_ +  \,r^{\,\beta _1 }  + {\rm B}_ -  \,r^{\,\beta _2 }
\,\frac{\Gamma ( - 2\beta _2  - 1)}{\Gamma ( - \beta _2
)}\,(2i\omega )^{2\beta _2  + 1}\,.
\eeq
As we hoped, the stretched far-field solution contains powers of $r$ that
exactly match the ones appearing in the stretched near-horizon solution (\ref{pffhe}).
Then, by matching also the corresponding multiplicative coefficients,
we find
\bea
&~& \hspace*{-1cm}\tilde B \equiv \frac{B_ - }{B_ + } = \frac{
\left[(1+a_*^2)\,r_h^{n+1}\right]^{2-2\beta-D_*}}{(2i\omega )^{2\beta _2 + 1}}\,
\times \nonumber \\[1mm] &~& \hspace*{2cm} \frac{
\Gamma(2\beta + D_*-2)\,\Gamma(2+\alpha -\beta - D_*)\,\Gamma(1+\alpha-\beta)\,
\Gamma( - \beta _2 )}{
\Gamma(\alpha+\beta + D_* -1)\,\Gamma(\alpha+\beta)\,
\Gamma(2-2\beta - D_*)\,\Gamma (-2\beta _2  - 1)}\,. \label{tildeBeq}
\eea
We remind the reader that the coefficients $D_*$, $\alpha$ and $\beta$ are given
in Eqs. (\ref{Dstar}), (\ref{alpha}) and (\ref{beta}), respectively, while
the coefficients $\beta_{1,2}$ are defined in Eq. (\ref{pffhe}).

In order to finally compute the absorption probability, we need first to expand
the far-field solution (\ref{heff}) in the limit  $r \rightarrow \infty$. Then, we
find \cite{Abramowitz}
\bea
P_{FF} &\simeq&  \frac{e^{ - i\omega r} }{r}\left[B_+\,\frac{\Gamma (2 + 2\beta _1 )}
{\Gamma (1 + \beta _1 )}\,\frac{e^{i\pi (\beta _1  + 1)}}{(2i\omega )^{\beta _1+1}}
+B_-\,(2i\omega )^{\beta _2} \right] +
\frac{e^{i\omega r} }{r}\frac{B_+\,\Gamma (2 + 2\beta _1)}
{\Gamma (1 + \beta _1 )\,(2i\omega )^{\beta _1 +1}} + ... \nonumber \\[1mm]
&\equiv& A^{(\infty)}_{in}\,\frac{e^{ - i\omega r}}{r } +
A^{(\infty)}_{out}\,\frac{e^{i\omega r} }{r}\,.
\eea
As we see, the far-field solution (\ref{heff}) also reduces to a spherical
free-wave solution in the asymptotic infinity. We may thus use once again
the standard definition for the absorption probability to determine its
value
\beq \left|{\cal A}_{l,m}\right|^2  = 1 - \left|\frac{A_{out}^{(\infty)}}
{A_{in}^{(\infty)}} \right|^2 =
1 - \left|\frac{\Gamma (2  + 2\beta _1 )}{\Gamma (2+ 2\beta _1 )\,
e^{i\pi (\beta _1 +1 )} + \tilde B\,\Gamma(\beta_1+1)\,
(2i\omega )^{\beta _1+\beta _2 + 1}} \right|^2\,.\label{a2}
\eeq
The above expression, combined with Eq. (\ref{tildeBeq}), gives the absorption
probability for scalar fields, valid only in the high-energy regime, but with
no restrictions on the value of the angular momentum parameter $a$ apart from
the upper bound of Eq. (\ref{amax}). The corresponding absorption cross-section,
valid in the high-energy regime, then follows by using the formula
$\sigma_{l,m}=\pi |{\cal A}_{l,m}|^2/\omega^2$ and Eq. (\ref{a2}).

\begin{figure}
  \begin{center}
  \mbox{\includegraphics[width = 0.5 \textwidth] {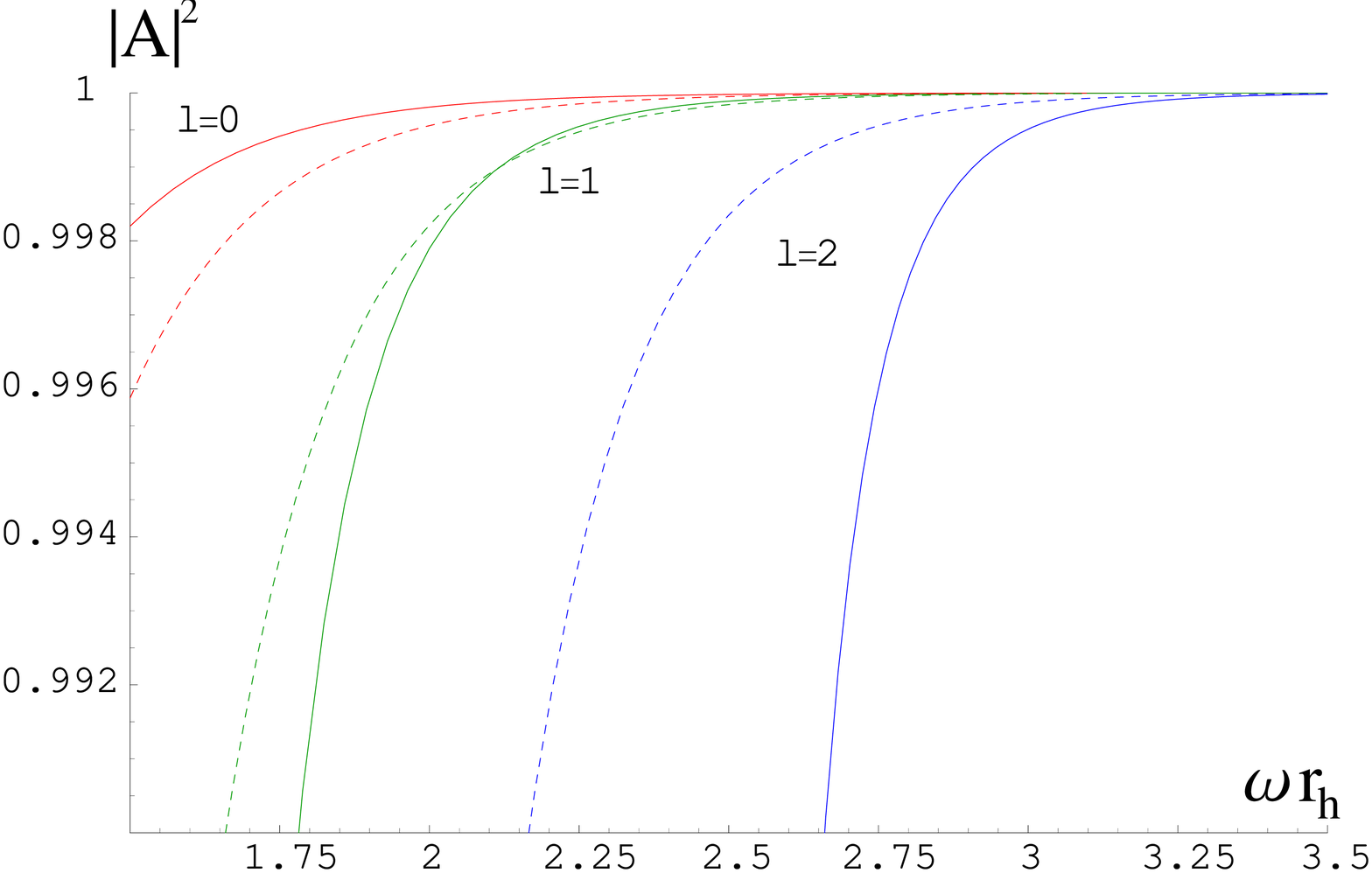}}
\hspace*{-0.5cm}
{\includegraphics[width = 0.5 \textwidth] {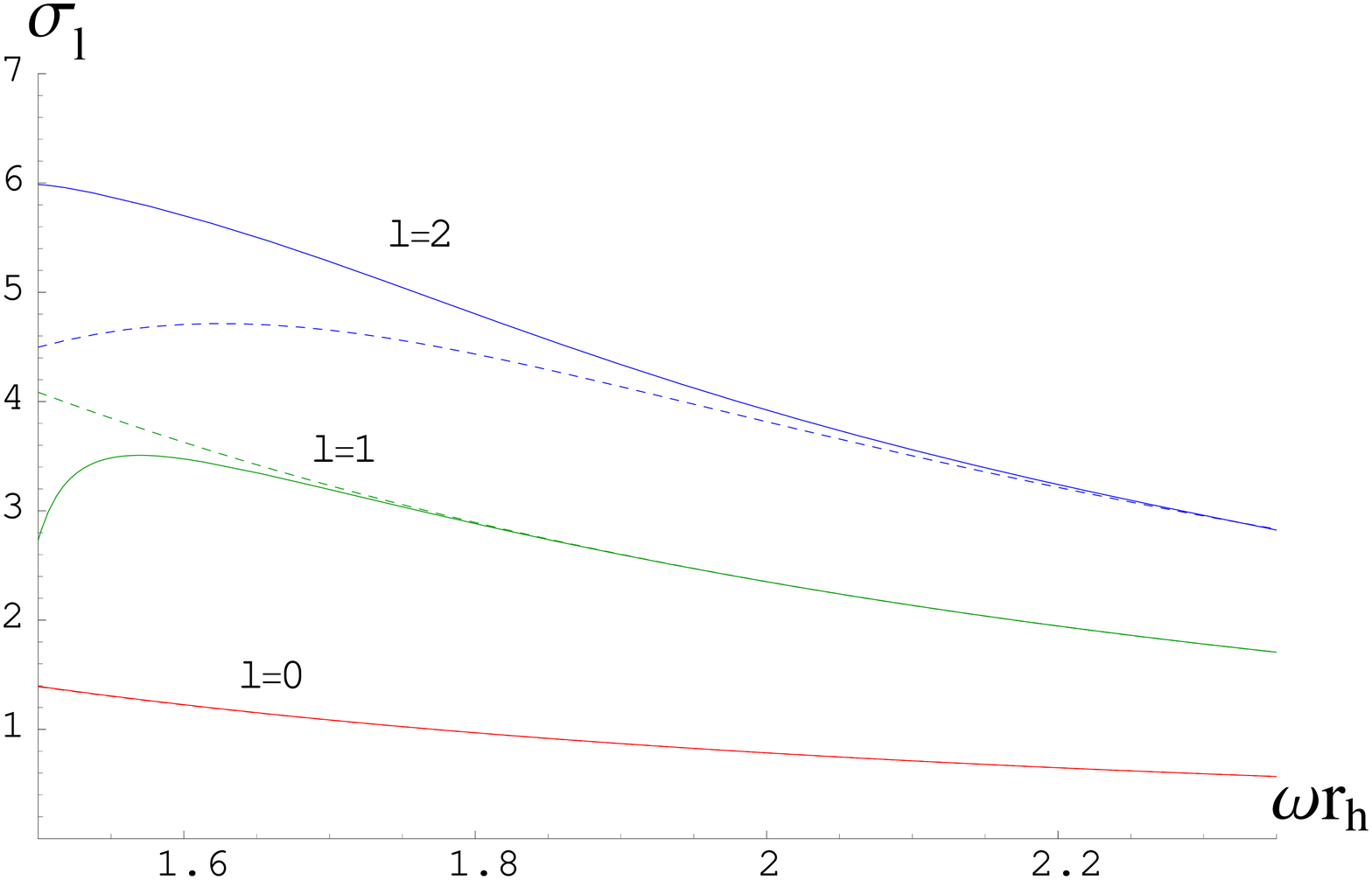}}
    \caption{{\bf (a)} Absorption probability $|{\cal A}_{l}|^2$ for brane scalar
   particles, for the modes $l=0,1,2$ and $m=0$, for $n=2$ and $a_*=0.3$,
   and {\bf (b)} absorption cross-section $\sigma_{l}$ (in units of $r_h^2$),
    for the modes $l=0,1,2$ (summed over $m$), and for the same values of $n$
    and $a_*$ as before. Again, the solid lines correspond to our analytic
    results, and the dashed lines to the exact numerical ones.}
    \label{fig-Kummer}
  \end{center}
\end{figure}

In Figs. \ref{fig-Kummer}(a,b), we depict the absorption probability $|{\cal A}_{l,m}|^2$
and cross-section $\sigma_{l,m}$, respectively, for brane scalar particles, in the
high-energy regime. Once again, the solid lines correspond to our analytic results,
following from Eq. (\ref{a2}), and the dashed lines to the exact numerical ones.
Figure \ref{fig-Kummer}(a) depicts the absorption probability for the indicative
case of the three lowest partial modes. We notice that our analytic results
match the exact numerical ones for large enough value of the energy $\omega$.
The lower the
value $l$ of the partial mode, the sooner the two results coincide. As the
energy parameter increases, the absorption probability quickly tends to unity --
as expected, highly energetic particles always overcome the gravitational barrier
outside the horizon of the black hole. This asymptotic behaviour, although
successfully reproduced by our analytic results, does not allow us to appreciate
the agreement between the two sets of results. For this reason, in
Fig. \ref{fig-Kummer}(b), we plot the absorption cross-section for the three
lowest partial waves summed over $m$: in the high-energy regime,
$\sigma_{l,m} \sim 1/\omega^2$ for all modes, and the asymptotic regime is
significantly more extended. The agreement between our analytic and numerical
results is now much clearer: for the mode $l=0$, the two results completely
coincide, while, for $l=1,2$, the exact matching is achieved at gradually larger
values of energy. Once matched, the two results remain identical as $\omega$
increases further until the zero asymptotic value -- for the individual partial
modes -- is reached. Given the increased difficulty in integrating numerically
both the radial and angular part of the scalar equation of motion over an
extended energy regime (for example, see \cite{HK1, DHKW}), the above solution
could be used to analytically extrapolate a numerical solution to arbitrarily
large values of the energy parameter $\omega$.


\subsection{High-Energy Asymptotic and Geometrical Optics Limits}

As the energy of the particle emitted from a black hole increases,
the total absorption cross-section $\sigma_{\rm abs}=\sum_{l,m}
\sigma_{l,m}$ reaches a high-energy asymptotic value, in an
oscillatory way. Although each partial cross-section
$\sigma_{l,m}$ asymptotes zero at the high-energy regime, the
superposition of an infinite number of partial waves, each one
reaching its maximum value at a gradually larger value of $\omega$
as $l$ increases, creates this constant asymptotic value. This
asymptotic limit has been studied in the past for a Schwarzschild
black hole, both in the four-dimensional
\cite{Misner:1974qy,Page:1976df,Sanchez:1976xm,MacGibbon:1990zk}
and $(4+n)$-dimensional case \cite{HK1,Emparan:2000rs}. For a
rotating black hole, the corresponding study was performed in 4
dimensions in \cite{Frolov:1998wf}, and in 5 dimensions in
\cite{Jung-rot}.

Here, we will attempt to give a comprehensive study of the high-energy
asymptotic limit of the total absorption cross-section for scalar fields
living on the brane-induced line-element of a $(4+n)$-dimensional rotating
black hole. As we will see, similarly to the case of a Schwarzschild-like
induced-on-the-brane line-element, the number of transverse dimensions,
although inaccessible to the brane-localised scalar fields, affects the
value of the high-energy asymptotic limit of the absorption cross-section.
The value of the angular momentum parameter $a_*$ of the higher-dimensional
black hole will also be found to have an effect on the value of $\sigma_{\rm abs}$.
Although our analytic results describe successfully, as we saw in previous
sections, both the low- and high-energy regimes of the absorption probability and
cross-section, no analytic solution currently exists that smoothly connects
the two solutions over the intermediate energy regime. The emergence of the
high-energy asymptotic limit of the total absorption cross-section strongly relies
on the contribution of the low-energy regime (where the $l=0$ mode dominates),
the intermediate-energy regime (where all modes have a significant contribution)
and the high-energy regime (where higher modes dominate). As a result, in order
to accurately derive the high-energy asymptotic value of $\sigma_{\rm abs}$, and in the
absence of a global analytic solution, the use of exact numerical analysis
is imperative.

\begin{figure}
  \begin{center}
  {\includegraphics[width = 0.7 \textwidth] {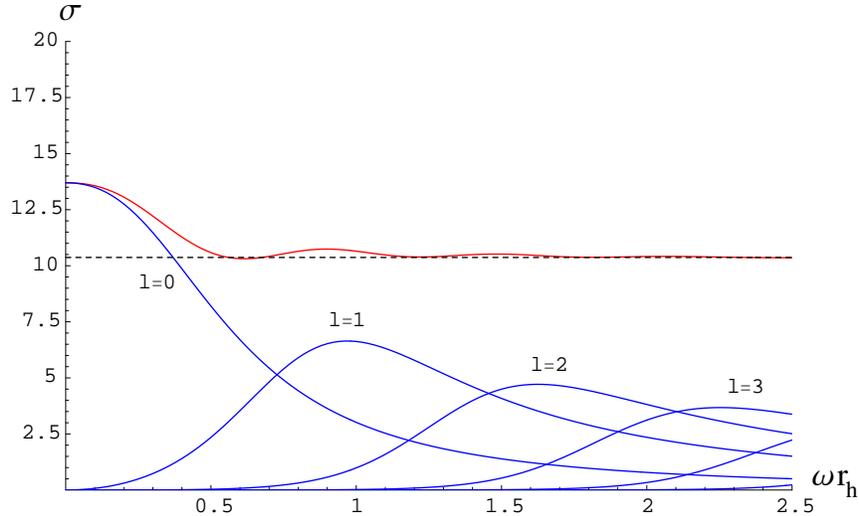}}
    \caption{Partial absorption cross-sections $\sigma_{l}$,
     for the modes $l=0,1,2,3,4$ (lower set of curves), and the total absorption
    cross-section $\sigma_{\rm abs}$ (upper curve) for $n=2$ and $a_*=0.3$,
     both in units of $r_h^2$. The
   dashed line denotes the value obtained by using the geometrical optics limit
   (\ref{min-par}).}
    \label{cross-exact}
  \end{center}
\end{figure}

In Fig. \ref{cross-exact}, we therefore present exact numerical results for the
absorption cross-section $\sigma_l=\sum_m \sigma_{l,m}$ for the partial modes
$l=0,1,2,3,4$, as well as for the total absorption cross-section $\sigma_{\rm abs}$
for a brane-localised scalar field, for the indicative case of $n=2$ and
$a_*=0.3$. The emergence of a constant high-energy asymptotic value for the
total cross-section is obvious. In the context of our analysis, we have studied
the behaviour of $\sigma_{\rm abs}$ for a range of values of $n$ and $a_*$,
with our results being displayed in Table 1. Note that, while for low values
of $n$ and $a_*$, a relatively small number of partial modes needs to be summed
over (for the case depicted in Fig. \ref{cross-exact}, only modes up to $l=5$ were
included in the calculation), as either $a_*$ or $n$ takes large values, an
increasing number of partial waves needs to be taken into account.
In addition, as either $a_*$ or $n$ increases further, the asymptotic value
of $\sigma_{\rm abs}$ emerges at continuously larger values of $\omega$, which
significantly increases the computing time.

From the entries of Table 1, one may observe the strong dependence of
the high-energy asymptotic limit of the absorption cross-section on
both the number of transverse-to-the-brane dimensions and the angular
momentum of the black hole. As in the non-rotating case \cite{HK1},
$\sigma_{\rm abs}$ is strongly suppressed as $n$ increases. On the other
hand, an increase in the value of $a_*$ causes an enhancement in the
value of $\sigma_{\rm abs}$. For $a_*=0$, the values of $\sigma_{\rm abs}$
match, as expected, the ones obtained for a scalar particle in a
Schwarzschild-like projected brane background \cite{HK1}. We would
also like to note that a feature that seemed to hold in the
5-dimensional case \cite{Jung-rot}, namely that the high-energy asymptotic
value of $\sigma_{\rm abs}$ is close to the low-energy one, disappears for
general $n$. This can be clearly seen in Fig. \ref{cross-exact}, or by
comparing the entries of Table 1 with the value of $\sigma_0$, from Eq.
(\ref{lowsigma}), that dominates the low-energy value of $\sigma_{\rm abs}$:
for general $n$, the two sets of values are distinctly different.

\begin{table}[t]
\begin{center}
$\begin{array}{|c|cccccc|}  \hline \hline
{\rule[-1mm]{0mm}{5.5mm}
\hspace*{0.4cm} {\bf a_* \backslash n} \hspace*{0.5cm}} &
\hspace*{0.5cm} 1 \hspace*{0.5cm} & \hspace*{0.5cm} 2 \hspace*{0.5cm} &
\hspace*{0.5cm} 3 \hspace*{0.5cm} & \hspace*{0.5cm} 4 \hspace*{0.5cm} &
\hspace*{0.5cm} 5 \hspace*{0.5cm} & \hspace*{0.5cm} 6 \hspace*{0.5cm} \\ \hline
{\rule[0mm]{0mm}{4.5mm} 0.0} & 12.6 & 9.6  & 8.2 & 7.3  & 6.7 & 6.2\\
0.3 & 13.6 & 10.4 & 8.6 & 7.6  & 7.0 & 6.5\\
0.5 & 15.7 & 11.5 & 9.5 & 8.4  & 7.6 & 7.1\\
0.7 & 18.7 & 13.2 & 10.7& 9.4  & 8.5 & 7.9\\
1.0 & 25.1 & 16.6 & 13.2& 11.4 & 10.3 & 9.5\\
1.5 & 40.7 & 24.1 & 18.6& 15.8 & 14.0 & 12.9\\
2.0 & 62.8 & 33.6 & 25.2& 21.1 & 18.7 & 17.2\\
\hline \hline
\end{array}$
\end{center}
\caption{High-energy asymptotic values of the total absorption cross-section
$\sigma_{\rm abs}$, in units of $r_h^2$, as a function of $n$ and $a_*$.}
\end{table}

In the case of a non-rotating black hole, the geometrical optics
limit has been successfully used to explain the high-energy
asymptotic value of the absorption cross-section $\sigma_{\rm
abs}$ both in the pure 4-dimensional case
\cite{Misner:1974qy,Page:1976df,Sanchez:1976xm,MacGibbon:1990zk}
and $(4+n)$-dimensional one \cite{HK1,Emparan:2000rs}. In the
higher-dimensional case, for particles living on the brane, the
geometrical optics analysis showed that the Schwarzschild black
hole behaves as a perfect absorber of a radius given by
\beq r_c = r_h \left(\frac{n + 3}{2} \right)^{\frac{1}{n + 1}}
\sqrt {\frac{n + 3}{n + 1}}\,.\label{rcnd}
\eeq
The absorption cross-section is then given by the target area,
$\sigma_{\rm abs}=\pi r_c^2$. The values following from this expression,
for different $n$, are in perfect agreement with the numerical
ones found in \cite{HK1}, and displayed here in the first row of Table 1.

Here, we will attempt to perform a similar study, in an axially-symmetric
black-hole brane background, and investigate the potential connection between
the analytic values that follow from this analysis and the exact numerical
ones depicted in Table 1, for $a_* \neq 0$. For this, we will closely follow the
method described in \cite{Frolov:1998wf}. Although that formalism was developed
for the case of a pure 4-dimensional Kerr black hole, it holds identically
for the case of a projected-on-the-brane rotating black hole, with the only
difference appearing in the exact expression of the metric function
$\Delta(r)$. We therefore present here only the basic assumptions and
the final equation that describe the particle's trajectory.

The line-element (\ref{induced}), in which the brane-localised particles
propagate, is invariant under translations of the form $t\rightarrow
t+\Delta t$ and $\phi\rightarrow\phi+\Delta\phi$. The corresponding Killing
vectors $\xi^\mu_{(t)}=\delta_t^\mu$ and $\xi^\mu_{(\varphi)}=\delta_\varphi^\mu$
then lead to the conserved conjugate momenta $p_t\equiv-E$ and $p_\phi\equiv-L_z$.
The brane metric also possesses a Killing tensor $\xi_{\mu\nu}$, that leads to an
additional conserved quantity ${\cal Q}= \xi_{\mu\nu}\,p^\mu p^\nu -(Ea+L_z)^2$.
Combining the above, the equation of motion of a particle with rest mass $m$,
i.e. $p_\mu\,p^\mu =m^2$, takes the form \cite{Frolov:1998wf}
\beq
\Sigma \,\frac{dr}{d\lambda} =  \pm {\cal{R}} ^{1/2}\,, \quad
{\cal{R}} = \left[ {E(r^2  + a^2 ) + L_z a} \right]^2  - \Delta
\left[ {m^2 r^2  + (L_z  + aE)^{\rm{2}}  + {\cal{Q}}}\right]\,,
\label{one}
\eeq
where $\lambda$ is the affine parameter of the trajectory. The
conserved quantity ${\cal Q}$ takes the explicit form
$ {\cal{Q}} = L^2  - L_z ^2  - a^2 (E^2  - m^2 )\cos ^2\theta_\infty$,
where $\theta_\infty$ is the value of the azimuthal angle as the particle
approaches the black hole from infinity, and $L$ the total angular
momentum of the particle.

A particle approaching a rotating black hole from infinity, may do so
by following a number of possible trajectories. 
Here, we will be interested in the case of a massless particle with its
trajectory being either transverse ($\theta_\infty=\pi/2$) or parallel
($\theta_\infty=0,\pi$) to the rotation axis. Starting with the first case,
we notice that, for motion strictly on the equatorial plane,
$\cos\theta_\infty=0$ and $L=L_z$. Then, ${\cal Q}=0$, and
Eq. (\ref{one}) takes the form
\beq
\left(\Sigma\,\frac{dr}{d\lambda}\right)^2 = E^2\left[
b^2 (a^2-\Delta) +2 b\,\frac{a \mu}{r^{n-1}} + (r^2+a^2)^2 -a^2 \Delta\right].
\label{two}
\eeq
In the above, we have defined $L/E \equiv b$, where $b>0$ is the impact
parameter of the particle. For the above equation to be consistent, its
right-hand-side, or equivalently the expression inside the square brackets,
must be positive-definite. Since the particle approaches the black hole
from large $r$, we focus our attention on the radial regime outside the
ergosphere, where the coefficient of $b^2$, $(a^2-\Delta)$, can be shown
to be negative. Then, the constraint on the values of $b$ takes the form
$b_2<b<b_1$, where $b_{1,2}$ are the roots of the equation, following by
setting the right-hand-side of Eq. (\ref{two}) equal to zero.
However, it may easily be seen that $b_2<0$, therefore, the classically
acceptable regime is defined by the constraint  $0<b<b_1$. Particles with
impact parameters in this regime can access all values of the radial
coordinate, and thus reach the black-hole horizon, too, where they get absorbed.
According to the geometrical optics argument, then, the closest distance the
particle can get to the black hole without being captured is
\beq
r_c = \min (b_1)=\min \left(\frac{a\mu + r^{n + 1}
\sqrt {a^2  + r^2  - \frac{\mu}{r^{n - 1}}}}{r^{n + 1}-\mu}\right)\,.
\label{min}
\eeq
As a consistency check, we observe that, for $a=0$, the above expression reduces to
$r_c= \min \left(r/\sqrt {1 - \frac{\mu}{r^{n + 1}}}\right)$, which
leads directly to the result (\ref{rcnd}) for a non-rotating brane black-hole background
derived in \cite{Emparan:2000rs, HK1}. By further setting $n=0$, the purely
4-dimensional Schwarzschild case \cite{Misner:1974qy,Page:1976df,Sanchez:1976xm}
is also recovered, with $r_c=3\sqrt{3}\,r_h/2$.

\begin{table}[t]
\begin{center}
$\begin{array}{|c|cccccc|}  \hline \hline
{\rule[-1mm]{0mm}{5.5mm}
\hspace*{0.4cm} {\bf a_* \backslash n} \hspace*{0.5cm}} &
\hspace*{0.5cm} 1 \hspace*{0.5cm} & \hspace*{0.5cm} 2 \hspace*{0.5cm} &
\hspace*{0.5cm} 3 \hspace*{0.5cm} & \hspace*{0.5cm} 4 \hspace*{0.5cm} &
\hspace*{0.5cm} 5 \hspace*{0.5cm} & \hspace*{0.5cm} 6 \hspace*{0.5cm} \\ \hline
{\rule[0mm]{0mm}{4.5mm} 0.0} & 12.6 & 9.6  & 8.2 & 7.3  & 6.6 & 6.2\\
0.3 & 17.9 & 12.8 & 10.4 & 9.0  & 8.1 & 7.4\\
0.5 & 23.5 & 15.9 & 12.6 & 10.7  & 9.5 & 8.7\\
0.7 & 31.0 & 20.0 & 15.4& 12.9  & 11.4 & 10.3\\
1.0 & 46.0 & 27.6 & 20.6& 17.0 & 14.8 & 13.4\\
1.5 & 81.9 & 44.3 & 31.7& 25.6 & 22.2 & 19.9\\
2.0 & 131.6 & 65.4 & 45.3& 36.3& 31.3 & 28.2\\
\hline \hline
\end{array}$\\[4mm]
$\begin{array}{|c|cccccc|}  \hline \hline
{\rule[-1mm]{0mm}{5.5mm}
\hspace*{0.4cm} {\bf a_* \backslash n} \hspace*{0.5cm}} &
\hspace*{0.5cm} 1 \hspace*{0.5cm} & \hspace*{0.5cm} 2 \hspace*{0.5cm} &
\hspace*{0.5cm} 3 \hspace*{0.5cm} & \hspace*{0.5cm} 4 \hspace*{0.5cm} &
\hspace*{0.5cm} 5 \hspace*{0.5cm} & \hspace*{0.5cm} 6 \hspace*{0.5cm} \\ \hline
{\rule[0mm]{0mm}{4.5mm} 0.0} & 12.6 & 9.6  & 8.2 & 7.3  & 6.6 & 6.2\\
0.3 & 10.0 & 8.2 & 7.3 & 6.6  & 6.2 & 5.9\\
0.5 & 9.5 & 8.1 & 7.3 & 6.8  & 6.4 & 6.1\\
0.7 & 9.5 & 8.4 & 7.7& 7.3  & 6.9 & 6.7\\
1.0 & 10.5 & 9.6 & 9.0& 8.6 & 8.4 & 8.1\\
1.5 & 13.4 & 13.2 & 12.7& 12.3 & 12.1 & 11.9\\
 2.0 & 19.2 & 18.5 & 18.1& 17.8& 17.5 & 17.4\\
\hline \hline
\end{array}$
\caption{Absorption cross-section $\sigma_{\rm abs}$ (in units of $r_h^2$)
for particles moving in the equatorial plane of the axially-symmetric brane
black hole (\ref{induced}), for $a>0$ (upper sub-table) and $a<0$
(lower sub-table).}
\end{center}
\end{table}

For general $n$ and $a$, an analytic expression for the minimum distance $r_c$
is difficult to find. Nevertheless, a simple numerical analysis may lead
to the value of $r_c$ in units of $r_h$, after using Eq. (\ref{horizon}) to
eliminate the mass parameter $\mu$ from Eq. (\ref{min}). Then, through the
relation $\sigma_{\rm abs}=\pi r_c^2$, the corresponding absorption
cross-section may be found; its values, for
a variety of $n$ and $a_*$, are displayed in Table 2. The two sub-tables
correspond to the two possible orientations of the particle's angular
momentum $L$: as it approaches the black hole from infinity moving in the
equatorial plane, its angular momentum and the black-hole one can either
be parallel ($a L>0$) or anti-parallel ($a L <0$). Here, we have assumed
that $L>0$ always, and considered two different choices for the sign of the
angular momentum parameter of the black hole, $a>0$ and $a<0$, that correspond
to the first and second sub-table of Table 2, respectively. For $a<0$, the
sign of the $a\mu$-term in the numerator of Eq. (\ref{min}) is reversed,
a modification that leads to a lower value of $r_c$ and eventually of the
cross-section.

We now proceed to the case of a zero-mass particle coming from infinity in
an orbit parallel to the black hole's rotation axis. This translates into
$\cos^2\theta_\infty=1$ and $L_z=0$. In that case, we find
\beq
\left(\Sigma\,\frac{dr}{d\lambda}\right)^2 = E^2 (r^2 + a^2 )^2 - \Delta L^2.
\eeq
Defining, as before, $b\equiv L/E$, one may easily conclude that the above
equation is again consistent only if
\beq
b < \left( \frac{r^2  + a^2}{\sqrt \Delta} \right)\,.
\eeq
The above leads to the minimum distance of the particle's approach to the
black hole without being captured given by
\beq
r_c=\min \left(\frac{r^2  + a^2 }{\sqrt{a^2  + r^2  - \frac{\mu}{r^{n - 1}}}} \right)\,.
\label{min-par}
\eeq
For $a=0$, the above result also reduces to the Schwarzschild-like one (\ref{rcnd}),
as expected, since, in the absence of rotation, all directions of the particle's orbit
should give the same result. By using Eq. (\ref{min-par}), the values of the
corresponding absorption cross-section $\sigma_{\rm abs}$, in units of $r_h^2$, are
given in Table 3.

\begin{table}[t]
\begin{center}
$\begin{array}{|c|cccccc|}  \hline \hline
{\rule[-1mm]{0mm}{5.5mm}
\hspace*{0.4cm} {\bf a_* \backslash n} \hspace*{0.5cm}} &
\hspace*{0.5cm} 1 \hspace*{0.5cm} & \hspace*{0.5cm} 2 \hspace*{0.5cm} &
\hspace*{0.5cm} 3 \hspace*{0.5cm} & \hspace*{0.5cm} 4 \hspace*{0.5cm} &
\hspace*{0.5cm} 5 \hspace*{0.5cm} & \hspace*{0.5cm} 6 \hspace*{0.5cm} \\ \hline
{\rule[0mm]{0mm}{4.5mm} 0.0} & 12.6 & 9.6  & 8.2 & 7.3  & 6.6 & 6.2\\
0.3 & 13.7 & 10.4 & 8.7 & 7.8 & 7.1 & 6.6\\
0.5 & 15.7 & 11.6 & 9.7 & 8.6  & 7.9 & 7.3\\
0.7 & 18.7 & 13.5 & 11.2& 9.8  & 9.0 & 8.4\\
1.0 & 25.1 & 17.2 & 14.0& 12.3 & 11.3 & 10.5\\
1.5 & 40.9 & 25.7 & 20.5& 18.0 & 16.5 & 15.4\\
2.0 & 62.8 & 36.5 & 28.9& 25.3& 23.3 & 22.0\\
\hline \hline
\end{array}$
\end{center}
\caption{Absorption cross-section $\sigma_{\rm abs}$, in units of $r_h^2$,
for particles moving parallel to the rotation axis of the axially-symmetric
brane black hole (\ref{induced}).}
\end{table}

Let us now compare the various analytic values, obtained above by using
the geometrical optics limit, with the numerical ones for the high-energy
asymptotic value of the absorption cross-section. One should, of course,
be careful when a direct comparison of these results is made: the
values displayed in Tables 2 and 3 correspond to trajectories with
a specific azimuthal angle, while the numerical values of Table 1
are actually integrated over all angles. Nevertheless, a comparison
between the two sets of results could reveal, upon finding an
agreement, which type of trajectories may be used to account for
the value of the total cross-section more
accurately. A direct comparison of the entries of Tables
1 and 2 shows that the total cross-section is smaller than the
one corresponding to a trajectory lying on the equatorial plane
with $a L>0$, but larger than the one with $a L<0$. On the other
hand, by comparing the entries of Tables 1 and 3, we find that
there is an almost perfect agreement between these two sets of
results for low values of either $a_*$ or $n$. The same agreement
can be pictorially seen in Fig. \ref{cross-exact}, where the particular
case $n=2$ and $a_*=0.3$ is shown. We may thus conclude that
particle trajectories running parallel to the rotation axis of the black
hole lead to an absorption cross-section whose value is virtually identical
to the total one. As either $n$ or $a_*$ increases further, the
two sets deviate; in this parameter regime, the contribution of
all possible particle trajectories needs to be more carefully taken
into account before the value of the total absorption cross-section
can be justified.


\section{Conclusions}

The emission of Hawking radiation, i.e. the evaporation of a black
hole via the emission of elementary particles, takes place during the
{\it spin-down} and {\it Schwarzschild} phase of its life. Although
the emission during the Schwarzschild phase of a higher-dimensional
black hole was studied, both analytically and numerically, quite early,
the complexity of the gravitational background around a similar, but
rotating, black hole delayed the study of the spin-down phase.
During the last few years, numerical
studies have derived results for the various spectra characterising
the emission of elementary particles on the brane by a higher-dimensional
rotating black hole -- the most phenomenologically interesting emission
channel for the brane-localised observers. Nevertheless, no analytical
studies have been performed and no analytic expressions for the
fundamental quantities governing the emission of Hawking radiation,
such as the absorption probability, have ever been derived for an
arbitrary value of the number of extra dimensions $n$. In this
work, we have duly performed this task, and studied in
detail the properties of the absorption probability and absorption
cross-section for scalar fields emitted on the brane by the
$(4+n)$-dimensional axially-symmetric black hole.

As the complexity of the equation of motion, describing the propagation
of a scalar field in the axially-symmetric brane background, forbids the
derivation of a general solution for a particle with arbitrary frequency,
we were forced to focus our analysis on two particular energy regimes:
the low-energy one and the high-energy one. The low-energy regime was
studied in Section 3, where an analytic solution for the radial part
of the scalar-field wave-function was derived. This involved matching the
near-horizon and far-field asymptotic solutions in an intermediate regime,
and allowed calculation of the absorption probability. Our analytic
results, formally valid only for low values of the energy parameter
$\omega r_h$ and angular momentum parameter $a_*$, were compared
with the exact numerical results, that were also reproduced during
our analysis in an attempt to check the range of validity of our
approximations. The two sets of results were found to be in excellent
agreement in the low-energy regime, as expected. In addition, even for
moderately large values of $\omega r_h$ and $a_*$, the agreement
on both qualitative and quantitative levels still persisted.

The properties of the absorption probability in the low-energy regime,
as these follow from our analytic results, were then studied in detail.
Its dependence on the angular momentum numbers $(l,m)$ was investigated
first, with our analysis revealing the dominance of the lowest partial
mode $l=0$, and the one with $m=-|l|$, for fixed $l$. The absorption
probability was also found to strongly depend on the spacetime
topological parameters, namely the angular momentum parameter $a_*$
of the black hole and the number of extra dimensions $n$: the
non-superradiant modes with $m \leq 0$, were shown to be enhanced
with both $a_*$ and $n$, while the superradiant ones, with $m>0$,
were, on the contrary, suppressed -- however, outside the superradiant
regime, the latter modes were enhanced with $n$. Our analytic expression
for the absorption probability, valid in the low-energy regime, was then
expanded in the limit $\omega \rightarrow 0$, and the constant asymptotic
value of the absorption cross-section was derived. The analytic value was
shown to accurately reproduce, for small $a_*$, the exact numerical one,
that was equal to the area of the horizon of the projected-on-the-brane
axially-symmetric black hole, $4 \pi(a^2+r_h^2)$.

We subsequently turned our attention to the study of the scalar
equation of motion in the high-energy regime. By using an
analogous approximate method, and for the first time in the
literature, an analytic solution was derived that perfectly
matched the exact numerical one for large enough values of the
energy parameter $\omega r_h$. The value of $\omega r_h$, beyond
which the two solutions completely coincide, was shown to be
strongly mode-dependent: modes with small $l$ allowed the two
solutions to match fairly quickly, while, modes with larger $l$
had the matching taking place at an increasing value of
the energy parameter. By employing the exact numerical solution
for the absorption probability, valid at all energy regimes, we
were able to determine the constant asymptotic value of the total
absorption cross-section at the high-energy regime, and its
dependence on $n$ and $a_*$. Similarly to the
spherically-symmetric case, this asymptotic value is suppressed as
the number of extra dimensions increases. On the other hand, an
increase in the angular momentum of the black hole causes an
enhancement in the high-energy asymptotic value of $\sigma_{\rm
abs}$. A detailed analysis, based on the geometrical optics limit,
revealed that the asymptotic value of the absorption cross-section
in the high-energy limit is accurately reproduced by considering
particle trajectories approaching the black hole from infinity and
running parallel to the rotation axis of the black hole.

The analytic results, supplemented by exact numerical ones, derived in this
work, on the behaviour of the absorption probability and cross-section
for scalar particles propagating in an axially-symmetric brane black-hole
background, smoothly complement the sole previous analytic study of the
5-dimensional case \cite{IOP1}, as well as the numerical studies
of the Hawking radiation spectrum of \cite{HK2,DHKW, IOP2}. Given the
excellent agreement between our analytic solutions and the exact numerical
ones, in the low and high-energy regime, their use to derive the
corresponding emission rates would have led to results identical to
the ones already presented in the works cited above. For that reason,
we have refrained from performing this task here, and refer the interested
reader to those works. Instead, in this manuscript, we have focused our
attention on the derivation of closed-form expressions and study of the
properties of the absorption
probability and cross-section, that carry a significant amount of information
on particle properties as well as on properties of the spacetime. Apart
from their obvious theoretical interest, the above results may have a
phenomenological one too: both our analytic solutions could be reliably
used, in place of the exact numerical ones, to derive the energy emission
rates in the low and high-energy regimes; these could then be used for
the interpretation of any observable effects coming from an evaporating
black hole produced in a ground-based accelerator and centered in these two
frequency regimes. In the high-energy regime, in particular, where
constraints on the available running time may put limits on the derivation
of numerical data, our analytic solution may prove useful in removing
the need for numerical integration.

\bigskip

{\bf Acknowledgments.} We would like to thank C. Harris for
providing the basis for our numerical code, and E. Winstanley for
useful discussions in the early stages of this work. S.C and O.E.
acknowledge PPARC and I.K.Y. fellowships, respectively. The work
of P.K. is funded by the UK PPARC Research Grant PPA/A/S/
2002/00350. P.K. and K.T. acknowledge participation in the RTN
Universenet (MRTN-CT-2006035863-1). This research was co-funded by
the European Union in the framework of the Program $\Pi Y\Theta
A\Gamma O PA\Sigma-II$ of the {\textit{``Operational Program for
Education and Initial Vocational Training"}} ($E\Pi EAEK$) of the
3rd Community Support Framework of the Hellenic Ministry of
Education, funded by $25\%$ from national sources and by $75\%$
from the European Social Fund (ESF).


\end{document}